\newif\ifemulate
\emulatetrue

\ifemulate
   \documentclass{emulateapj}
   \slugcomment{Accepted for Publication in ApJ}
\else
   \documentclass[12pt, preprint]{aastex}
\fi

\lefthead{Rudnick et al.}
\righthead{Cosmic Mass Density Evolution}

\newcommand\lrest{${\rm L^{rest}_\lambda}$}

\renewcommand{\l}[1]{${\rm L}_{#1}^{rest}$}

\newcommand\bvr{$(B-V)_{rest}$}
\newcommand\ubr{$(U-B)_{rest}$}

\newcommand\uvr{$(U-V)_{rest}$}
\newcommand\zp{$z_{phot}$}
\newcommand\dzp{$\delta z_{phot}$}

\newcommand\Js{$J_s$}
\newcommand\Jsab{$J_{s,{\rm AB}}$}
\newcommand\Hab{$H_{{\rm AB}}$}
\newcommand\Ks{$K_{s}$}
\newcommand\Kstot{$K_{s}^{\rm tot}$}
\newcommand\Ksab{$K_{s,{\rm AB}}$}
\newcommand\Ksvega{$K_{s,vega}$}
\newcommand\Hvega{$H_{vega}$}
\newcommand\Jsvega{$J_{s,vega}$}

\newcommand\ang{${\rm \AA}$}
\newcommand\zs{$z_{spec}$}

\newcommand\lstar{${\rm L}^\star$}
\newcommand\lstarlocal{${\rm L}^\star_{\rm local}$}

\newcommand{\lstarlam}[1]{${\rm L}^{\star}_{#1}$}

\newcommand\mlstar{${\cal M_{\star}}/{\rm L}$}
\newcommand{\mlstarlam}[1]{${\cal M_{\star}}/{\rm L}_{#1}$}
\newcommand{\mlstarlamav}[1]{$\langle {\cal M_{\star}}/{\rm L}_{V}\rangle$}

\newcommand\msol{${\cal M_{\odot}}$}

\newcommand\jk{$(J_s - K_s)$}

\newcommand\uv{$(U-V)$}
\newcommand\lumsol{${\rm ~h_{70}^{-2}~L_{\odot}}$}
\newcommand\rhostar{$\rho_\star$}

\newcommand\jrest{$j^{\rm rest}_\lambda$}
\newcommand{\jrestlam}[1]{$j^{\rm rest}_{\rm #1}$}

\newcommand\magu{$U_{300}$}
\newcommand\magb{$B_{450}$}
\newcommand\magv{$V_{606}$}
\newcommand\magi{$I_{814}$}
\newcommand\magj{$J_{110}$}
\newcommand\magh{$H_{160}$}
\newcommand\maghtot{$H_{160}^{\rm tot}$}
\newcommand\magz{$z_{850}$}
\renewcommand\u{$U$}
\newcommand\ux{$UX$}
\newcommand\mb{$B$}

\newcommand\mv{$V$}
\newcommand\mr{$R$}
\newcommand\mi{$I$}
\newcommand\mj{$J$}
\newcommand\mh{$H$}
\newcommand\mk{$K$}
\newcommand{\muv}[1]{$ST-UV#1$}

\newcommand{\lthreshlam}[1]{${\rm L}_{#1}^{\rm thresh}$}
\newcommand{\taumod}[1]{$\tau_{\rm #1}$}
\newcommand{\mstar}{$\cal{M_{\star}}$}

\begin{document}
\title{Measuring the Average Evolution of Luminous Galaxies at $z<3$: The Rest-frame Optical Luminosity Density, Spectral Energy Distribution, and Stellar Mass Density \altaffilmark{1}}

\author{Gregory Rudnick\altaffilmark{2,3}
Ivo Labb\'{e}\altaffilmark{4,5}, Natascha M. F\"orster
Schreiber\altaffilmark{6}, Stijn Wuyts\altaffilmark{7}, Marijn
Franx\altaffilmark{7}, Kristian Finlator\altaffilmark{8}, Mariska
Kriek\altaffilmark{7}, Alan Moorwood\altaffilmark{9}, Hans-Walter
Rix\altaffilmark{10}, Huub R\"ottgering\altaffilmark{7}, Ignacio
Trujillo\altaffilmark{11}, Arjen van der Wel\altaffilmark{7}, Paul van
der Werf\altaffilmark{7}, \& Pieter G. van Dokkum\altaffilmark{12}}

\altaffiltext{1}{Based on observations with the
    NASA/ESA \textit{Hubble Space Telescope}, obtained at the Space
    Telescope Science Institute, which is operated by the AURA, Inc.,
    under NASA contract NAS5-26555.  Also based on observations
    collected at the European Southern Observatories on Paranal, Chile
    as part of the ESO program 164.O-0612}

\altaffiltext{2}{NOAO, 950 N. Cherry Ave, Tucson, AZ 85719, USA, \texttt{grudnick@noao.edu}}
\altaffiltext{3}{Leo Goldberg Fellow}
\altaffiltext{4}{Carnegie Observatories, 813 Santa Barbara Street, Pasadena, CA 91101, USA}
\altaffiltext{5}{Carnegie Fellow}
\altaffiltext{6}{Max-Planck-Institut f\"{u}r extraterrestrische Physik,  Giessenbachstrasse, Garching, D-85748, Germany}
\altaffiltext{7}{Leiden Observatory, PO BOX 9513, 2300 RA Leiden, Netherlands}
\altaffiltext{8}{Steward Observatory, 933 N. Cherry Av., Tucson, AZ, 85721}
\altaffiltext{9}{European Southern Observatory, Karl-Schwarzschild-Strasse 2, 85748 Garching, Germany}
\altaffiltext{10}{Max-Planck-Institut f\"{u}r Astronomie, K\"{o}nigstuhl 17, Heidelberg, D-69117, Germany}
\altaffiltext{11}{School of Physics \& Astronomy, University of Nottingham, University Park, Nottingham, NG7 2RD, UK}
\altaffiltext{12}{Astronomy Department, Yale University, P.O. Box 208101, New Haven, CT 06520-8101}

\begin{abstract}
 We present the evolution of the volume averaged properties of the
 rest-frame optically luminous galaxy population to $z\sim3$,
 determined from four disjoint deep fields with optical to
 near-infrared wavelength coverage.  Our use of independent lines of
 sight substantially mitigates field-to-field variations.  We select
 galaxies above a fixed rest-frame \mv-band luminosity of $3\times
 10^{10}$\lumsol\ and characterize their rest-frame ultraviolet
 through optical properties via the mean spectral energy distribution
 (SED).  To measure evolution we apply the selection criteria to a
 sample of galaxies from the Sloan Digital Sky Survey (SDSS) and
 COMBO-17 survey.  The mean rest-frame 2200\ang\ through \mv-band SED
 becomes steadily bluer with increasing redshift but at all redshifts
 $z<3$ the mean SED falls within the range defined by ``normal''
 galaxies in the nearby Universe.  Rest-frame 4000\AA/Balmer breaks
 are present in the volume averaged SED at all redshifts, indicating
 significant light and mass contributions from evolved populations,
 even at $z\sim 2.8$.  We measure stellar mass-to-light ratios
 (\mlstar) by fitting models to the rest-frame UV-optical SEDs.
 Multiplying our volume averaged \mlstarlam{V}\ estimates at each
 redshift by the measured mean \mv-band luminosity density, we derive
 the stellar mass densities \rhostar.  The stellar mass density in
 galaxies selected at a fixed luminosity has increased by a factor
 $3.5-7.9$ from $z=3$ to $z=0.1$, where the range includes the
 uncertainty due to field-to-field variance within our own data. If we
 use our observed \mlstarlam{V} evolution to select galaxies at a
 fixed mass, the stellar mass density evolves by a factor of $5.3-
 16.7$.  After correcting to total, the measured mass densities at
 $z<2$ lie below the integral of the star formation rate (SFR) density
 as a function of redshift as derived from UV selected samples after a
 standard correction for extinction.  This may indicate a systematic
 error in the total \rhostar\ or SFR($z$) estimates.  We find large
 discrepancies between recent model predictions for the evolution of
 \rhostar\ and our results, even when our observational selection is
 applied to the models.  Finally we determine that Distant Red
 Galaxies (selected to have $J_s - K_s>2.3$) in our \l{V}\ selected
 samples contribute $30\%$ and 64\% of the stellar mass budget at
 $z\sim2$ and $z\sim 2.8$ respectively.  These galaxies are largely
 absent from UV surveys and this result highlights the need for mass
 selection of high redshift galaxies.

\end{abstract}
\keywords{Evolution --- galaxies: formation --- galaxies: high redshift --- galaxies: stellar content --- galaxies: galaxies} 
\section{Introduction}
\label{Intro}

 Our knowledge of the high redshift Universe has increased rapidly
 over the past decade and is becoming ever more comprehensive.  The
 initial large advances in this field were enabled by the efficient
 selection of $z\sim3$ starforming galaxies based on their optical
 light, e.g. \citet{Stei96} and \citet{Stei99}, and by the development
 of efficient multi-object spectrographs on large 8-10m class
 telescopes.  Objects selected by these techniques allowed us for the
 first time to study statistically significant samples of galaxies at
 $z>1.5$.  Resultant follow-up work on these optically (rest-frame
 ultraviolet; UV) selected objects at $z\sim3$ (Lyman Break Galaxies;
 LBGs) and subsequent samples at $z\sim 2$ (BM/BX objects; Steidel et
 al. 2004; Adelberger et al. 2004) have demonstrated that these
 galaxies have generally low extinctions $E(B-V)\sim 0.15$ and modest
 ages ($<1$ Gyr) and stellar masses (\mstar$\sim 10^{10}$\msol) with a
 tail to high end values existing at bright \mk\ magnitudes,
 e.g. \citet{Saw97}, Papovich, Dickinson, \& Ferguson (2001; hereafter
 P01), \citet{Shap01}, \citet{Shap05}, \citet{Reddy05}.

 Although efficient in telescope time for observing objects with
 $R\lesssim25$, optical surveys will miss objects that have high
 extinctions or those with a lack of current vigorous star formation.
 In fact, most galaxies with \mstar$\gtrsim10^{11}$\msol\ have optical
 magnitudes too faint for optical spectroscopic followup
 \citep{Dokkum06}.  A comprehensive view of the high redshift Universe
 requires that UV selection be complemented with selection from light
 at least as red as the NIR, which corresponds to the rest-frame
 optical out to $z\sim 3$.  Deep NIR observations allow us to detect
 highly extincted (e.g. SCUBA galaxies; Smail et al. 1997) or passive
 galaxies at $z>1.5$ (e.g. Daddi et al. 2005), which have very little
 rest-frame UV emission.

 However, getting deep NIR data is observationally expensive and
 even over small fields requires substantial investments of 8-10m
 telescope time to reach depths fainter than \Ks$=20-21$.  Until the
 recent development of megapixel NIR arrays progress was measurable
 but slow, e.g. \citet{Hogg97} and \citet{Ber98}.  A large step
 forward was taken with the Faint Infrared Extragalactic Survey
 (FIRES; Franx et al. 2000; Rudnick et al. 2001; Labb\'e et al. 2003;
 F\"orster Schreiber et al. 2005) and with the NICMOS observations of
 the Hubble Deep Field North (HDF-N; P01; Dickinson et al. 2003).
 FIRES is an ESO large program which imaged in \Js, \mh, and \Ks\ the
 HDF-S and a field centered around the x-ray luminous cluster
 MS1054-03 at $z=0.83$.  At the time these fields had a unique
 combination of area and deep HST optical imaging and the ESO data
 extended this coverage to the \Ks-band.  The FIRES data were
 particularly valuable because the associated deep \Ks-band data
 provide access to the rest-frame \mv-band all the way out to $z\sim
 3$.  The very deep FIRES data allowed us to probe rest-frame $V$-band
 luminosities \l{V}\ down to $\sim 50\%$ of the local \lstarlam{V}\
 value.

 The powerful combination of very deep NIR and optical data now make
 it possible to assemble representative samples of the Universe at
 $z>1.5$.  For example, \citet{Franx03} and \citet{Dokkum03} have
 shown that galaxies selected to have \jk$>2.3$ are luminous in the
 rest-frame optical, likely have high \mstar, but would be almost
 completely missed by rest-frame UV selected surveys.  Subsequent
 studies of these Distant Red Galaxies (DRGs) have shown that they
 have systematically older ages, higher extinctions, higher \mstar,
 and comparable or even higher star formation rates than the optically
 selected LBGs at comparable redshifts and \mk\ magnitudes
 \citep{Dokkum04, Forster04, Labbe05, Pap05, Dokkum06} and may
 contribute significantly to the stellar mass density at $z>1.5$
 (Rudnick et al. 2003 -- hereafter R03).  Other author have also found
 galaxy populations that would be missed by rest-frame UV selection,
 highlighting the need for a comprehensive selection of galaxies
 (e.g. Daddi et al. 2004).


 While individual properties of galaxies are indeed important, for
 some quantities the modeling of the volume averaged population may
 yield more robust results.  For example R03 demonstrated that bursty
 SFHs cause a smaller systematic error in the mean stellar
 mass-to-light ratio \mlstar\ of the galaxy population - and hence
 estimates of \rhostar\ - than when it is determined from averages of
 \mlstar\ estimates of individual galaxies.  The utility of using
 cosmically averaged quantities has also been demonstrated by other
 authors across a range of topics.  Using the luminosity density
 determinations at many wavelengths \citet{Mad98} tried to constrain
 the shape of the cosmic SFH by fitting the different bands
 simultaneously.  \citet{Pei99} derived independent constraints on the
 cosmic SFH by modeling the cosmic gas content, metallicity, and
 extragalactic background light.  Using the FIRES data on the HDF-S
 R03 modeled the rest-frame optical mean colors and luminosity
 densities to derive the evolution in the stellar mass density at
 $z<3$.  At low redshifts \citet{Baldry02} and \citet{Gla03} modeled
 the mean galaxy spectrum from 2dF and the Sloan Digital Sky Survey
 (SDSS; York et al. 2000) to constrain the cosmic SFH.

 Using the deepest existing \Ks-band data R03 showed for the HDF-S
 that the evolution in the volume averaged rest-frame optical colors,
 e.g. \ubr, \bvr, and \uvr, at $z<3$ was monotonic, with bluer colors
 at higher redshifts.  The colors also lay close to the locus of
 colors occupied by individual normal galaxies in the local Universe
 \citep{Lar78, Jansen00a}.  They also found that the evolution in the
 rest-frame colors could be approximated by smooth SFHs with a
 constant metallicity and dust.  Models with these SFHs were then used
 to convert the \uvr\ to \mlstarlam{V}\ and hence to a mass density.

 A primary goal of this paper is to extend the analysis of R03.  One
 major area of improvement is that this paper will model the full
 rest-frame UV through optical volume averaged SED of luminous
 galaxies in the Universe, allowing more freedom in the choice of dust
 attenuation, SFH, and age.  The full rest-frame UV through optical
 information also gives us more insight as to the nature of the mean
 stellar population of luminous galaxies at high redshift.  Perhaps
 the most important improvement with respect to R03 is that we include
 data from four disjoint fields, covering a total area of $\approx
 98.8$ arcminutes.  This is crucial since the field-to-field
 variations in the number densities of galaxies are expected to be
 large in small fields.  In addition, multiple fields are important to
 characterize the mean contribution of different galaxy populations,
 e.g. DRGs, to the volume averaged SED and \rhostar.  For example the
 field-to-field variations in the number densities of DRGs is known to
 be high with large differences between the HDF-N and HDF-S (Labb\'e
 et al. 2003; hereafter L03).

 In this paper we combine FIRES data from the HDF-S and MS1054-03
 fields with those from the HDF-N catalog of P01 and D03 and a catalog
 from the Great Observatories Origins Deep Survey (GOODS) images of
 the Chandra Deep Field South (CDF-S; Wuyts et al. in preparation).
 Using photometric redshifts from these four catalogs we derive the
 rest-frame luminosities, luminosity densities and colors.  We examine
 the trends in color and redshift and compute the full rest-frame
 2200\ang\ through \mv-band volume averaged SED.  We then apply the same
 analysis to a galaxy sample from SDSS and from the COMBO-17 survey
 and use it to derive the evolution of mean SED properties in a
 consistent manner from $z\sim 3$ down to $z=0$.  Finally, we fit the
 full rest-frame mean SED with a set of simple models and use these to
 derive the mean \mlstarlam{V}\ and \rhostar\ in each redshift bin.

 In \S~\ref{data_sec} we describe the data.  In \S~\ref{methods_sec}
 we review the methods that we use to calculate photometric redshifts,
 rest-frame luminosities, luminosity densities, and global colors.  In
 \S~\ref{results_sec} we present the results including the evolution
 of the rest-frame luminosity density, color, and volume averaged SED.
 In this section we also present the model fits to the mean SED and
 the derived \mlstar\ and \rhostar\ evolution.  We discuss these results
 in \S~\ref{discuss_sec} and present our conclusions and summary in
 \S~\ref{sum_sec}.  Throughout this paper we assume
 $\Omega_\mathrm{M}=0.3,~\Omega_{\Lambda}=0.7,~\mathrm{and~H_o}=70~{\rm
 h_{70}~km~s^{-1} Mpc^{-1}}$ unless explicitly stated otherwise.

\section{The Data}
\label{data_sec}

 We combine results from four different fields: the HDF-S, MS1054-03,
 HDF-N, and CDF-S.  Below we briefly describe the data reduction and
 the construction of catalogs.  All magnitudes are quoted in the Vega
 system unless specifically noted otherwise.  Our adopted conversions
 from Vega system to the AB system are \Jsvega~= \Jsab~- 0.90,
 \Hvega~= \Hab~- 1.38, and \Ksvega~= \Ksab~- 1.86 \citep{BB88}.

\subsection{FIRES Data}

 The reduced images and photometric catalogs for the two FIRES fields
 are available online through the FIRES homepage at
 \\\texttt{http://www.strw.leidenuniv.nl/$\sim$fires}.

\subsubsection{The HDF-S}

 A complete description of the FIRES HDF-S observations, reduction
 procedures, and the construction of photometric catalogs is presented
 in detail in L03; we outline the important steps below.

 The FIRES data on the WFPC2 field of the HDF-S is comprised of 101.5
 hours of exposure with ISAAC \citep{Moor97} on the VLT.  These
 exposures were split into 33.6, 32.3, and 35.6 hours in \Js, \mh, and
 \Ks\ respectively.  The data were taken in service mode at the VLT and
 have a mean image quality better $0\farcs48$ in all bands.  These NIR
 data were combined with the very deep optical data from WFPC2 taken
 by \citet{Cas00}.

 Objects were detected in the \Ks-band image with version 2.2.2 of the
 SExtractor software \citep{Ber96}.  For consistent photometry between
 the space and ground-based data, all images were then convolved to
 $0\farcs48$, corresponding to the effective resolution of the NIR
 band with the worst average seeing.  Photometry was then performed in
 the \magu, \magb, \magv, \magi, \Js, \mh, and \Ks-band images using
 specially tailored isophotal apertures defined from the detection
 image.  In addition, a measurement of the total flux in the \Ks-band,
 \Kstot, was obtained using an aperture based on the SExtractor
 \textit{AUTO} aperture, which includes a conservative aperture
 correction.  The effective area of the HDF-S is 4.74 square
 arcminutes, including only areas of the chip with exposure time
 $\geq20\%$ of the total integration time in all bands.  The accuracy
 of photometric redshifts (to be discussed in \S\ref{photz_sec}) and
 rest-frame optical luminosities in our \Ks-selected sample is very
 dependent on the quality of the NIR data (e.g. Rudnick et al. 2001)
 and once the S/N in the NIR bands becomes too low the quality of the
 photometric redshifts becomes too poor for a useful analysis.  Since
 the \Ks-band is our shallowest NIR band a cut there is a conservative
 proxy for a cut in the S/N of the other NIR bands, keeping in mind
 the range of galaxy NIR colors.  The final catalog has 358 objects
 with \Kstot$<23.14$, which for point sources corresponds to a
 10$\sigma$ S/N in the custom isophotal aperture.

\subsubsection{The MS1054-03 Field}

 The MS1054-03 observations, data reduction, and catalog construction
 are described in detail in \citet{Forster06}.  This field has an
 x-ray detected cluster at $z=0.83$ and at the time of the FIRES
 proposal was the field with the best combination of depth and area
 observed with WFPC2.  The WFPC2 data in \magv\ and \magi\ are presented
 in \citet{Dokkum00}.  The FIRES observations of this field consisted
 of 78 hours of \Js, \mh, and \Ks\ imaging with ISAAC, supplemented
 with $UBV$ imaging with FORS1 on the VLT.  The observations were
 split over four pointings.  The total effective area is 23.5 square
 arcminutes, including only those pixels with exposure time $\geq20\%$
 of the total integration time in all bands.  Observing conditions in
 the MS1054-03 field were generally similar to the HDF-S and since the
 exposures were split over 4 pointings the depth is 0.7 magnitudes
 shallower than in the HDF-S.

 Object detection and catalog construction were performed in an
 identical way as in the HDF-S.  The final catalog has 1380 sources
 with \Kstot$<22.34$, which also corresponds roughly to a 10$\sigma$
 detection for point sources.

\subsection{HDF-N}

 The WFPC2 data on the HDF-N are presented in \citet{Wil96}.  The data
 reduction of the NICMOS \magj\ and \magh\ data and the Kitt Peak \Ks\
 data is described by \citet{Dick99} and \citet{Dick00} as is the catalog
 construction\footnote{The reduced images and photometric catalogs are
 available from Mark Dickinson, \texttt{mdickinson@noao.edu}}.
 Objects were detected in a weighted sum of the \magj\ and \magh\
 images, which are individually much deeper than the \Ks-band image.
 The HST \magu, \magb, \magv, \magi, \magj, and \magh\ data were then
 convolved to the same PSF and the fluxes were measured in matched
 apertures.  The ground-based \Ks\ data, which has much worse image
 quality, was extracted using the ``TFIT'' method described in P01 to
 achieve consistent colors between the space-based and ground-based
 images.  A total magnitude in the \magh-band was estimated using the
 SExtractor \textit{AUTO} aperture.  The area of the HDF-N is 5 square
 arcminutes.  The final catalog has 854 sources with
 \maghtot(AB)$<26.5$.

 The HDF-N is unique among our fields in that the the detection is not
 done in the \mk-band.  Nonetheless, the \magh-band data are
 sufficiently deep so that we are complete to the rest-frame \mv-band
 luminosity limit above which we select galaxies for this study (see
 \S~\ref{ldens_meas_sec}).

\subsection{CDF-S}

 From GOODS/EIS observations of the CDF-S (data release version 1.0) a
 \Ks-band selected photometric catalog is constructed, described by
 Wuyts et al. (in preparation).  ISAAC imaging on the VLT provides 72
 hours of exposure in \mj, 55 in \mh\ and 122 in \Ks, split over 21, 12
 and 23 pointings respectively.  GOODS zeropoints were adopted for \mj\
 and \Ks.  The \mh-band zeropoint was obtained by matching the stellar
 locus on a $J-K$ vs $J-H$ color-color diagram to the stellar locus in
 HDF-S and MS1054.  The difference with the official GOODS \mh-band
 zeropoint varies across the CDF-S but on average our \mh-band
 zeropoints is $\approx 0.1$ magnitudes brighter.  ACS imaging
 provides photometry in \magb, \magv, \magi, and \magz\ bands.  All
 images were smoothed to match the NIR pointing with the worst seeing,
 $0\farcs64$.  Identical procedures as in the HDF-S and MS1054 fields
 were applied to detect objects and construct the catalog.  A total
 effective area of 65.6 square arcminutes is well exposed in all
 bands.  The final catalog contains 1588 objects with \Kstot$< 21.34$
 in this area.  At \Kstot$=21.34$ the median S/N in the \Ks\ isophotal
 aperture is $\sim 12$.

\subsection{DRG selection}
 Because the depth across our four fields is non-uniform we do not
 consistently apply a magnitude selection for DRGs.  This may produce
 a slight error when comparing the DRGs to non-DRGs in each field.  In
 addition, DRGs have different magnitude distributions in the
 different fields \citep{Forster04} and this may slightly bias the
 mass budgets determined in \S\ref{massbudget_sec}.  In all fields our
 limiting \Ks\ magnitude for DRG selection was dictated by the depth of
 our \Js\ data, which must be deep enough so that non-detections in \Js\
 can still be selected as DRGs.

 In all fields DRGs were selected to have $z_{phot}>1.5$.  In the
 HDF-S 12 DRGs were selected to have $J_s - K_s > 2.3$, \Kstot$<22.5$
 and $z_{phot}>1.5$.  In the MS1054-03 field 31 DRGS were selected at
 \Kstot$<21.7$.  The brighter \Ks\ limit in comparison to the HDF-S is
 a direct result of the shallower data.  In the CDF-S 65 DRGs were
 selected with \Kstot$<21.14$.  This is 0.2 magnitudes shallower than
 the CDF-S catalog depth to insure reliable colors for objects that
 are red in \jk.

 We select DRGs in the HDF-N to have \Kstot$<21.05$, which corresponds to
 a 5-$\sigma$ detection in $K_s$.  The very deep NICMOS \magj\
 observations allow robust color measurements of red objects at this
 \Ks\ limit.  The selection of DRGs in the HDF-N is complicated,
 however, by the use of the \magj\ filter, which is significantly bluer
 (effective wavelength of $1.13\mu m$ for \magj\ and $1.25\mu m$ for
 \Js) and about twice as broad (c.f. the NICMOS Instrument Handbook).
 It is therefore not appropriate to select DRGs using a
 $J_{110}-K_s>2.3$ color cut.  Instead we compute a synthetic
 $J_s-K_s$ color using our best fit photometric redshift template (see
 \S\ref{photz_sec}) to derive the $J_s$ magnitude.  To assess the
 reliability of this method we first synthesize the $J_{110}-K_s$
 color and compare it the observed color.  At \Kstot$<21.05$ the outlier
 resistant normalized absolute median deviation ($\sigma_{NMAD}$;
 equal to the rms for a Gaussian distribution) in the colors is 0.05
 for all sources and 0.14 for those sources at $1.6<z<3.5$.  Computing
 the synthetic $J_s - K_s$ color for every galaxy we find two DRGs in
 the HDF-N with $1.6<z<3.5$.

\section{Methods}
\label{methods_sec}

\subsection{Photometric Redshifts and Rest-Frame Luminosities}
\label{photz_sec}

 Photometric redshifts \zp\ for all galaxies are derived using an
 identical code to that presented in \citet{Rud01} and R03 but with a
 slightly modified template set.  This code models the observed SED
 using non-negative linear combinations of a set of 8 galaxy
 templates.  As in R03 we use the E, Sbc, Scd, and Im SEDs from
 \citet{CWW80}, the two least reddened starburst templates from
 \citet{Kinn96} and a synthetic template corresponding to a 10 Myr
 year old simple stellar population (SSP) with a \citet{Sal55} stellar
 initial mass function (IMF).  In this paper we have added a 1 Gyr old
 SSP with a Salpeter IMF.

 Using spectroscopic redshifts \zs\ we have determined that
 $\sigma_{NMAD}(z_{spec} - z_{phot}~/~(1+z_{spec})=0.06$, 0.06, and
 0.08 at $z<4$, $0<z<1.5$, and $1.5\leq z < 4$ respectively.

 It is important to remember, however, that the spectroscopic samples
 in the HDF-S and HDF-N are highly biased toward UV bright objects at
 high redshift, e.g. LBGs.  In the MS1054-03 and CDF-S field
 spectroscopy has been performed on \Ks\ selected sources and those
 selected to be red in $J_s-K_s$ and in $I-H$.  The photometric
 redshift accuracy for DRGs is $\sigma_{NMAD}=0.11$ and 0.09 in in
 MS1054 and the CDF-S respectively.  There are, however, very few DRGs
 with secure \zs\ measurements (4 in MS1054-03 and 3 in the CDF-S)
 causing the determination of the \zp\ accuracy itself to be uncertain
 at this point.  From these first determinations, the accuracy of the
 \zp\ measurements is still adequate for the construction of the
 luminosity densities and colors that we use here.

 The two sided 68\% confidence intervals on \zp, i.e. \dzp, are
 computed using the monte carlo simulation described in R03.  In
 general the differences between \zp\ and \zs\ are well reflected by
 \dzp.  \dzp\ can therefore be used to judge the reliability of the \zp\
 estimates.

 Rest-frame luminosities are computed using the method described in
 the appendix of R03.  We compute rest-frame luminosities \lrest\ in
 two UV bands centered approximately at 2200 and 2700\ang.  These
 filters correspond to the \muv{22} and \muv{27} filters from (Bruzual
 \& Charlot 2003; hereafter BC03).  In addition we compute rest-frame
 luminosities in the \ux, \mb, \mv, \mr, and \mi\ filters of
 \citet{Bess90}.  For these filters we use $M_{\odot,U}=+5.66$,
 $M_{\odot,B}=+5.47$, $M_{\odot,V}=+4.82$, $M_{\odot,U}=+4.28$, and
 $M_{\odot,U}=+3.94$.  For each galaxy we only compute \lrest\ in
 rest-frame filters that are at shorter observed wavelengths than the
 \Ks\ filter.  In all cases where a spectroscopic redshift is available
 we compute the rest-frame luminosities fixed at \zs.

\subsubsection{Star Identification}

 Stars in all four fields were identified by spectroscopy, by fitting
 the object SEDs with stellar templates from \citet{Haus99}, and by
 inspecting their morphologies, as in R03.  In well exposed regions of
 the HDF-S, MS1054-03, HDF-N, and CDF-S data we identified 29, 68, 16,
 236 stars respectively down to \Kstot$<23.14$, \Kstot$<22.34$,
 \maghtot(AB)$<26.5$, and \Kstot$<21.34$.  All stars were excluded
 from the analysis

\subsection{Computing Luminosity Densities and Mean Colors}
\label{ldens_meas_sec}

\begin{figure}
\ifemulate 
	\epsscale{1.2}
\else
	\epsscale{1.0}
\fi
\plotone{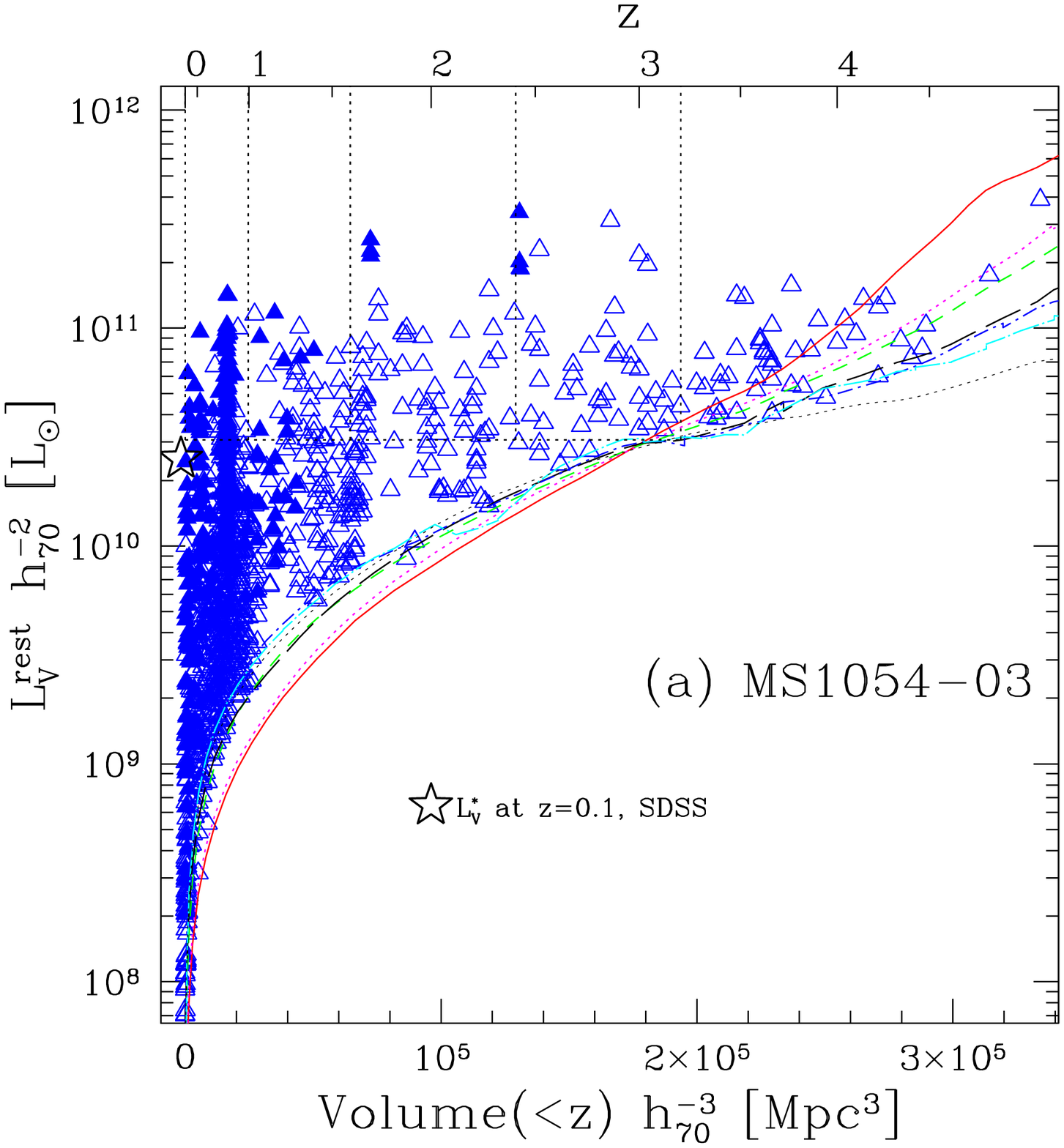}
\caption {The distribution of rest-frame \mv\
 luminosities as a function of enclosed co-moving volume and $z$ is
 shown for MS1054-03, the HDF-S, the HDF-N, and the CDF-S in figures
 (a), (b), (c), and (d) respectively for galaxies with \Kstot$<22.34$,
 \Kstot$<23.14$, \maghtot(AB)$<26.5$, and \Kstot$<21.34$.  Galaxies which have
 spectroscopic redshifts are represented by solid points and for these
 objects \lrest\ is measured at \zs.  The large star in each panel
 indicate the value of \lstarlocal\ from \citet{Blanton03b}.  The
 tracks represent the values of \lrest\ for seven template spectra
 normalized at each redshift to the limiting magnitude in that field.
 The specific tracks correspond to the E (solid), Sbc (dot), Scd
 (short dash), Im (long dash), SB1 (dot--short dash), SB2 (dot--long
 dash), and 10my (dot) templates.  The horizontal dotted line each
 panel indicates the luminosity threshold \lthreshlam{V} above which
 we measure the rest-frame luminosity density \jrest\ and the vertical
 dotted lines in each panel mark the redshift boundaries of the
 regions for which we measure \jrest.}
\label{lumvol_fig}
\end{figure}

 In Figure~\ref{lumvol_fig} we show the \l{V} vs. cumulative enclosed
 co-moving volume and redshift for each field.  The tracks in each
 plot give the \l{V} limit for a set of galaxy templates that
 corresponds to the observed 90\% completeness limit in each image.
 From this plot we choose the minimum \l{V} for which we are highly
 complete to the largest redshift over the largest area.  The CDF-S
 image has the largest area but the data in this field are also our
 shallowest.  Choosing an \l{V} limit such that we are complete out to
 $z<3$ in the CDF-S would cause us to throw away many galaxies with
 useful photometry in the other fields.  Instead we compromise by
 choosing a threshold such that we are complete out to $z<3$ for the
 HDF-S, HDF-N, and MS1054, while still being complete out to $z<2.41$
 in the CDF-S.  This \l{V} limit is \lthreshlam{V}~$>3\times
 10^{10}$\lumsol\ corresponding to the \Kstot$=22.34$ magnitude limit
 of the MS1054-03 field.  All computations of mean properties
 presented hereafter are computed for galaxies with
 \l{V}$>$\lthreshlam{V}.  For the CDF-S, therefore, average quantities
 are only computed at $z<2.41$.

 We compute the luminosity density \jrest\ in every rest-frame filter
 in four bins which span the redshift ranges $0<z\leq1.0$, $1.0<z\leq
 1.6$, $1.6<z\leq2.41$, and $2.41<z\leq3.2$ and which are centered at
 redshifts of 0.73, 1.33, 2.01, and 2.8 respectively.  These centers
 were chosen to split the co-moving volume in each bin evenly.  The
 latter two bins have equal co-moving volumes.  The 0.73 and 1.33 bins
 take this same volume and split it approximately 40/60\%.  This was
 done partly to sample better the large amount of time spanned by the
 $z=0-1.6$ interval and partly because of the rich cluster at $z=0.83$
 in the MS1054-03 field.  We don't wish to bias our analysis by this
 rich cluster and by splitting the lowest redshift bin we exclude the
 lowest redshift bin (containing the cluster) in the MS1054-03 field
 without throwing out the valuable $1.0<z<1.6$ information in that
 field.  We compute \jrest\ in an identical fashion as described in
 R03, by adding up the luminosities of all galaxies in each bin with
 \l{V}$>$\lthreshlam{V}.  \footnote{The MS1054-03 galaxies at $z>0.83$
 are lensed by the foreground cluster.  We correct for the effects of
 lensing using the method described in
 Appendix~\ref{lum_dens_lens_app} using the weak lensing map of this
 cluster by \citet{Hoekstra00}.  Using the redshift distribution of
 sources the inferred magnifications range from 5-20\%.} A more
 sophisticated method such as a $V/V_{max}$ method is not needed since
 our threshold is in rest-frame luminosity and was chosen such that we
 are complete at all redshifts.  As in R03 we exclude galaxies with
 photometric redshift confidence intervals $\delta z / (1 + z) > 0.16$
 but correct for their contribution to \jrest\ under the assumption
 that they have the same luminosity distribution as those sources with
 small redshift errors.  Uncertainties in \jrest\ are computed via a
 bootstrapping method in which 1000 samples are drawn from the
 subsample of galaxies with \l{V}$>$\lthreshlam{V}, with replacement
 allowed.  The size of each bootstrap sample is a Poisson distributed
 number drawn from the measured number of galaxies with
 \l{V}$>$\lthreshlam{V}.  For each iteration we recompute the
 correction for galaxies with large photometric redshift confidence
 intervals and compute the luminosity density.  We store the bootstrap
 iterations of the luminosity density in every band and use them later
 when performing the model fits in \S~\ref{modfit_sec}.

 It is important to note here that these \jrest\ values are lower
 limits to the total \jrest, since we do not extrapolate the
 luminosity function (LF) below \lthreshlam{V}.  We choose this method
 because the faint end slope of the rest-frame optical LF is
 observationally unconstrained at high redshift, for example, even our
 extremely deep \Ks-band data in the HDF-S reach to only $\sim50\%$
 of the present day \lstar.  To circumvent this limitation, many
 authors adopt a faint end slope from lower redshift bins and apply
 this to the higher redshift data.  We prefer to avoid the uncertain
 extrapolation to the faint end of the LF by restricting ourselves to
 the observed data.

 To compute the luminosity densities averaged over the different
 fields we combine them weighting by their respective volumes,
 i.e. the total solid angle of each field.  The mean colors we
 generate from the \jrest\ values, e.g.
\begin{equation}
(U-V)_{rest} = -2.5~\times~{\rm log}_{10}~j^{\rm rest}_{U} + M_{\odot, U} 
+ 2.5~\times~{\rm log}_{10}~j^{\rm rest}_{V} - M_{\odot, V}.
\label{coleq}
\end{equation}
 In such a formalism the mean color corresponds to the luminosity
 weighted mean colors of all galaxies with \l{V}~$>$~\lthreshlam{V}.
 As in R03 we corrected the \ubr\ for the \bvr\ dependent contribution
 of emission lines, as determined using the spectra of the Nearby
 Field Galaxy Survey (NFGS; Jansen et al. 2000b). 

 We point out that the measured values are weighted heavily to the
 largest field, which at $z<2.4$ is the CDF-S.  Because of the small
 number of fields and their very different weighting the determination
 of the field-to-field variance in our mean value of \jrest\ is not
 well statistically defined.

\begin{figure}
\ifemulate 
	\epsscale{1.2}
\else
	\epsscale{1.0}
\fi
\plotone{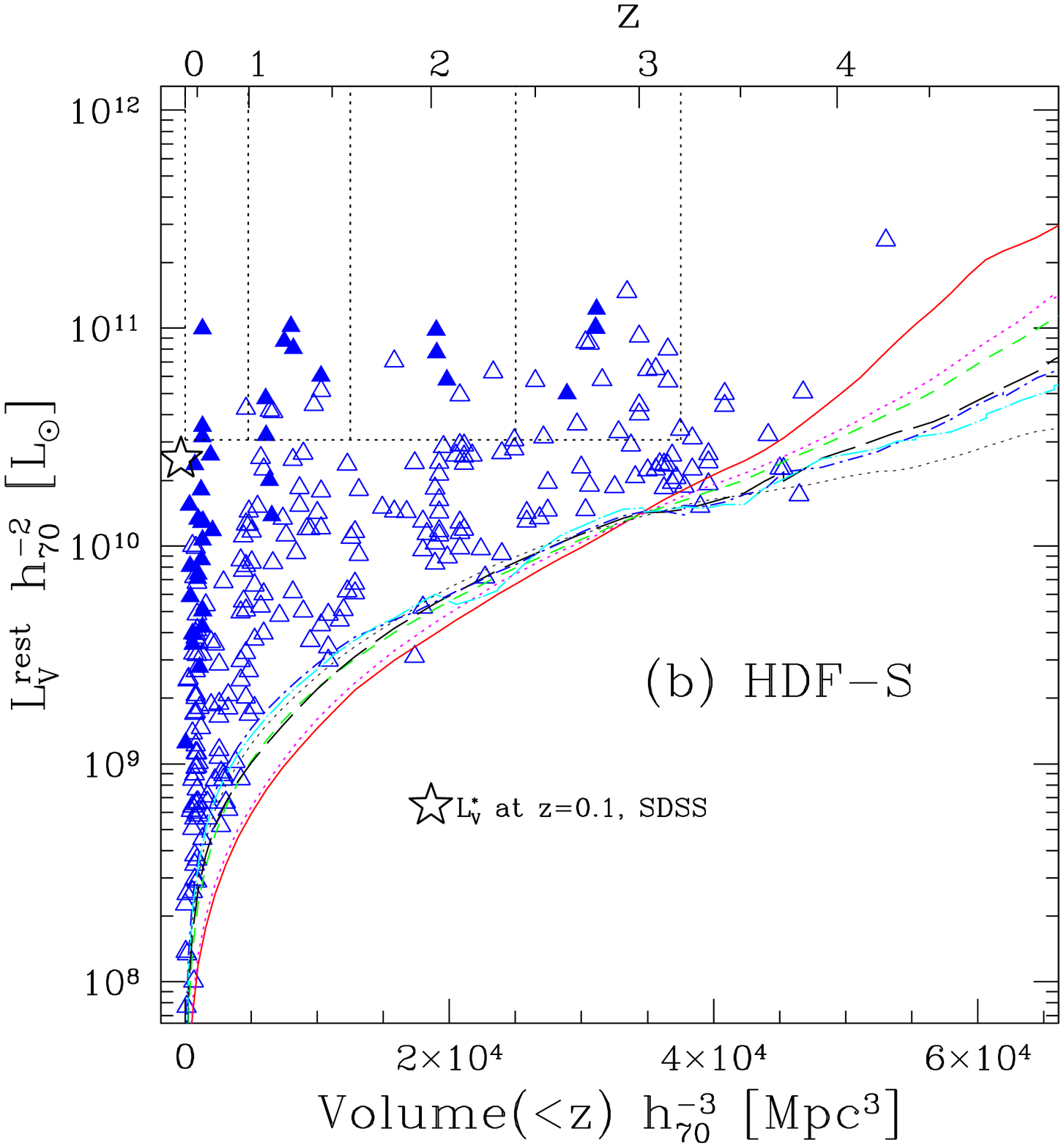}
\end{figure}

\begin{figure}
\ifemulate 
	\epsscale{1.2}
\else
	\epsscale{1.0}
\fi
\plotone{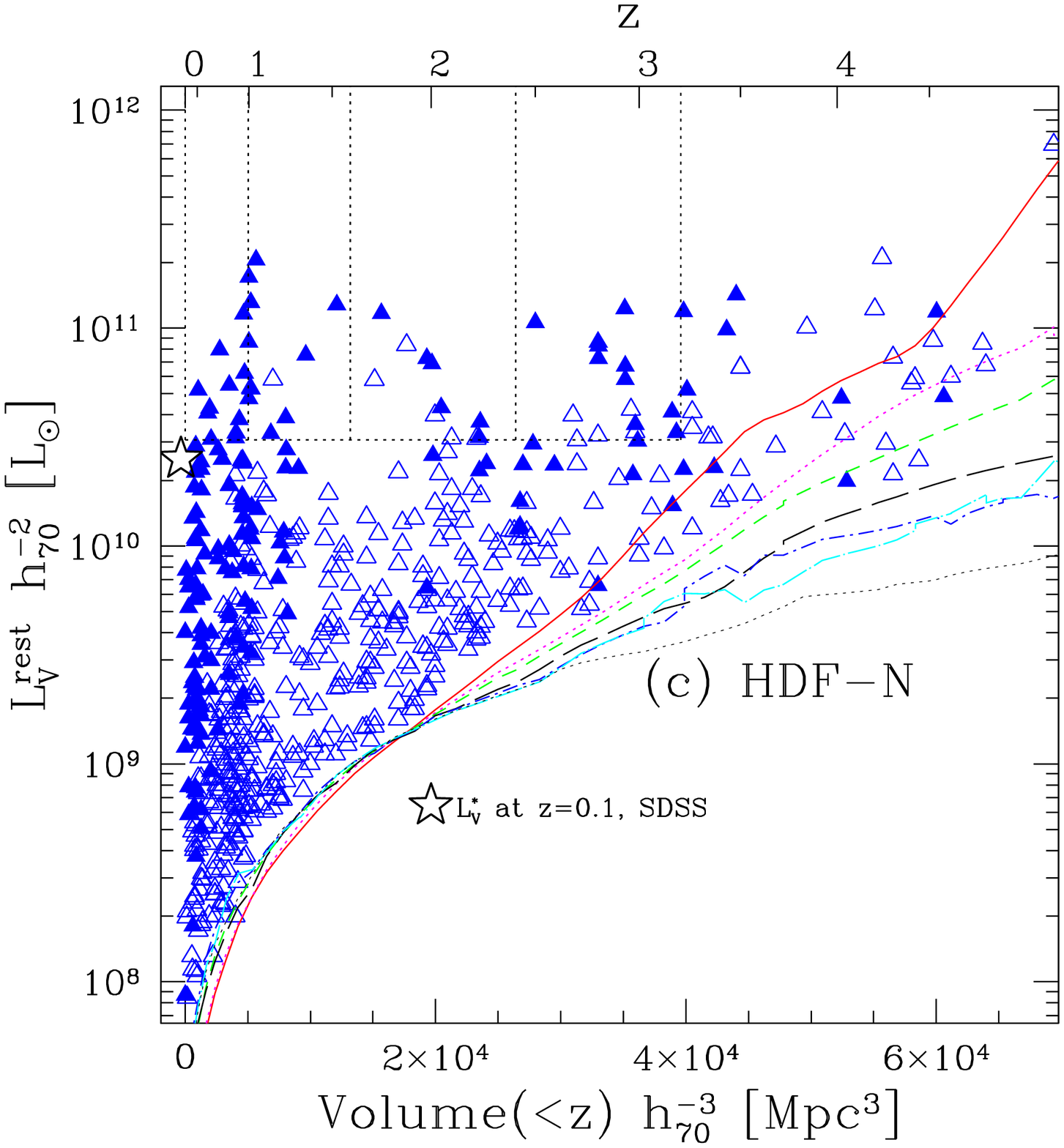}
\end{figure}

\begin{figure}
\ifemulate 
	\epsscale{1.2}
\else
	\epsscale{1.0}
\fi
\plotone{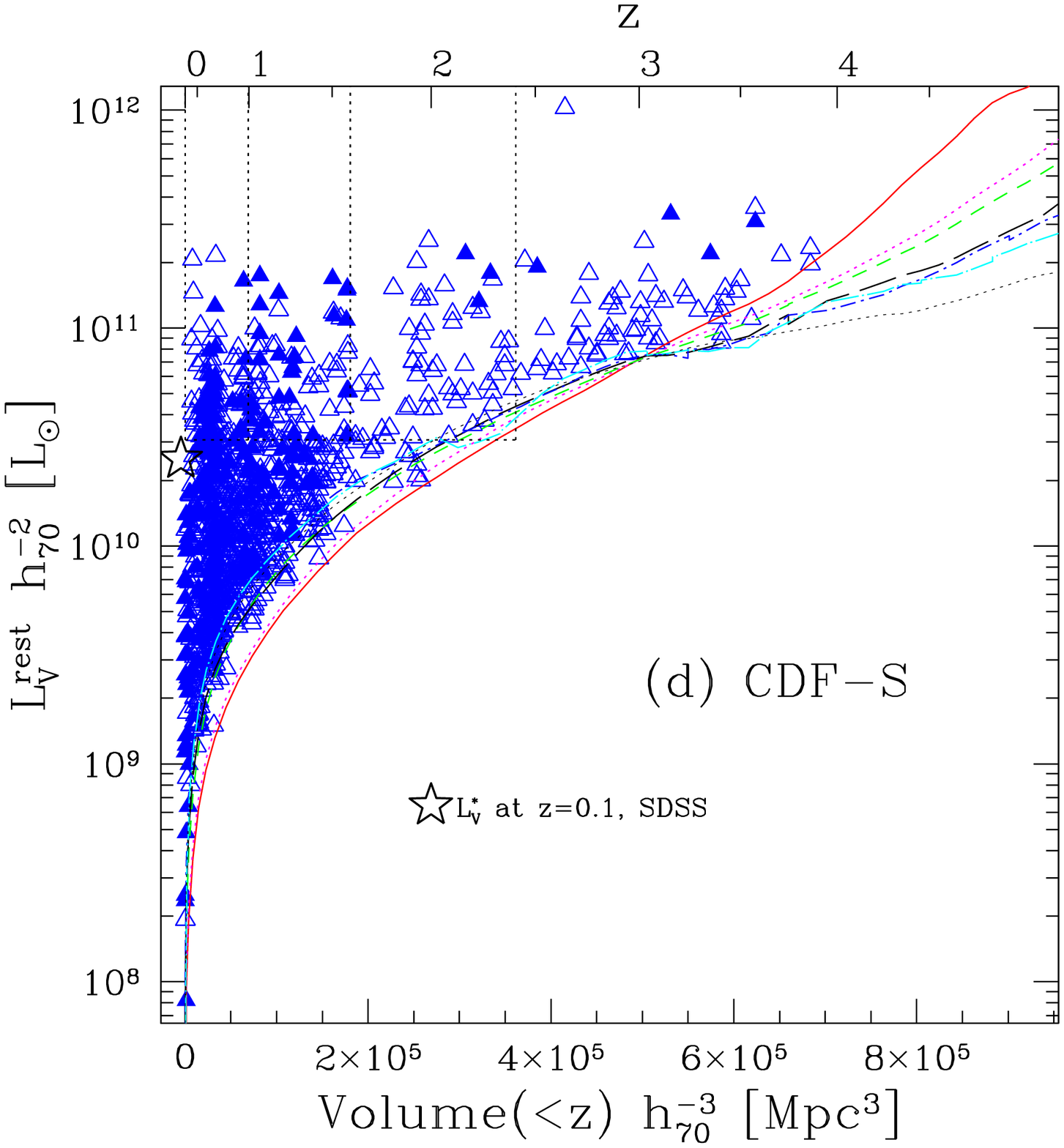}
\end{figure}

\subsubsection{Construction of Low Redshift Comparison Samples}
\label{lowz_sec}

 All deep surveys are limited by a small volume at $z\lesssim 1$,
 e.g. even the GOODS CDF-S field encloses only $6.6\times 10^4{\rm
 Mpc}^3$ or a cube of $40$Mpc on a side, 50\% of the distance to the
 Coma cluster. To build up a large volume at low redshift it is
 necessary to include a large redshift range in a single bin, which at
 low redshift smears together large epochs of time.  To supplement our
 data at these low redshifts we utilized data from the COMBO-17 survey
 and from SDSS.

 We computed \jrest\ at $z<0.2$ for the SDSS Early Data Release (EDR;
 Stoughton et al. 2002) as in R03, deriving \jrest\ in $UBVRI$ from
 maps of number density as a function of \lrest\ and rest-frame color.
 These maps were computed as described in \citet{Blanton03a}.  While
 the Poisson errors in the SDSS are negligible, cosmic variance and
 systematic errors do contribute to the uncertainties.  For a more
 conservative error estimate, we adopt 10\% errors on the luminosity
 density.  For the SDSS luminosity function, our \lthreshlam{V}
 encompasses 28\% of the total light.

 To provide data at $0.2<z<0.9$ we used \jrest\ measurements from the
 COMBO-17 survey \citep{Wolf03}, which has a $\sim 30$ times larger
 survey area than the four deep fields combined.  Specifically we used
 a catalog with redshifts of 29471 galaxies at $z<0.9$, of which 7441
 had \l{v}$>$\lthreshlam{V} \footnote{The J2003c catalog; available at
 \texttt{http://www.mpia.de/COMBO/combo\_index.html}.}.  Using this
 catalog we calculated \jrest\ identically to the deep fields.  We
 divided the data into redshift bins of $\Delta z=0.2$ and counted the
 light from all galaxies contained within each bin which had
 \l{v}$>$\lthreshlam{V}.  The size of the redshift slices was chosen
 to well sample the $0.2<z<1$ redshift interval while still being
 large enough to include many sources in each bin.  The large solid
 angle of the COMBO-17 survey (0.75 deg$^2$) assured that these slices
 still contained considerable co-moving volume.  The formal 68\%
 confidence limits were calculated via a bootstrapping method as in
 \S~\ref{ldens_meas_sec}.  In addition, in Figure~\ref{lumdens_fig} we
 indicate the rms field-to-field variations among the three COMBO-17
 fields.  As also pointed out in \citet{Wolf03}, the field-to-field
 variations dominate the error in the COMBO-17 \jrest\ determinations.

\section{Results}
\label{results_sec}

 In this section we first present our estimates of \jrestlam{V} and of
 the evolution of the volume averaged color of luminous galaxies.  We
 then discuss the color evolution in terms of simple models.  We also
 present the full rest-frame UV to optical volume averaged SED of
 luminous galaxies and show its evolution to higher redshifts,
 highlighting the mean SEDs of different subpopulations.  Finally, we
 fit these mean SEDs with models and use the results to constrain the
 evolution in the global \mlstar\ and \rhostar, also highlighting the
 contribution to the mass budget by different galaxy subpopulations.

\subsection{The Luminosity Density}
\label{ldens_sec}

\begin{figure}
\ifemulate 
	\epsscale{1.2}
\else
	\epsscale{1.0}
\fi
\plotone{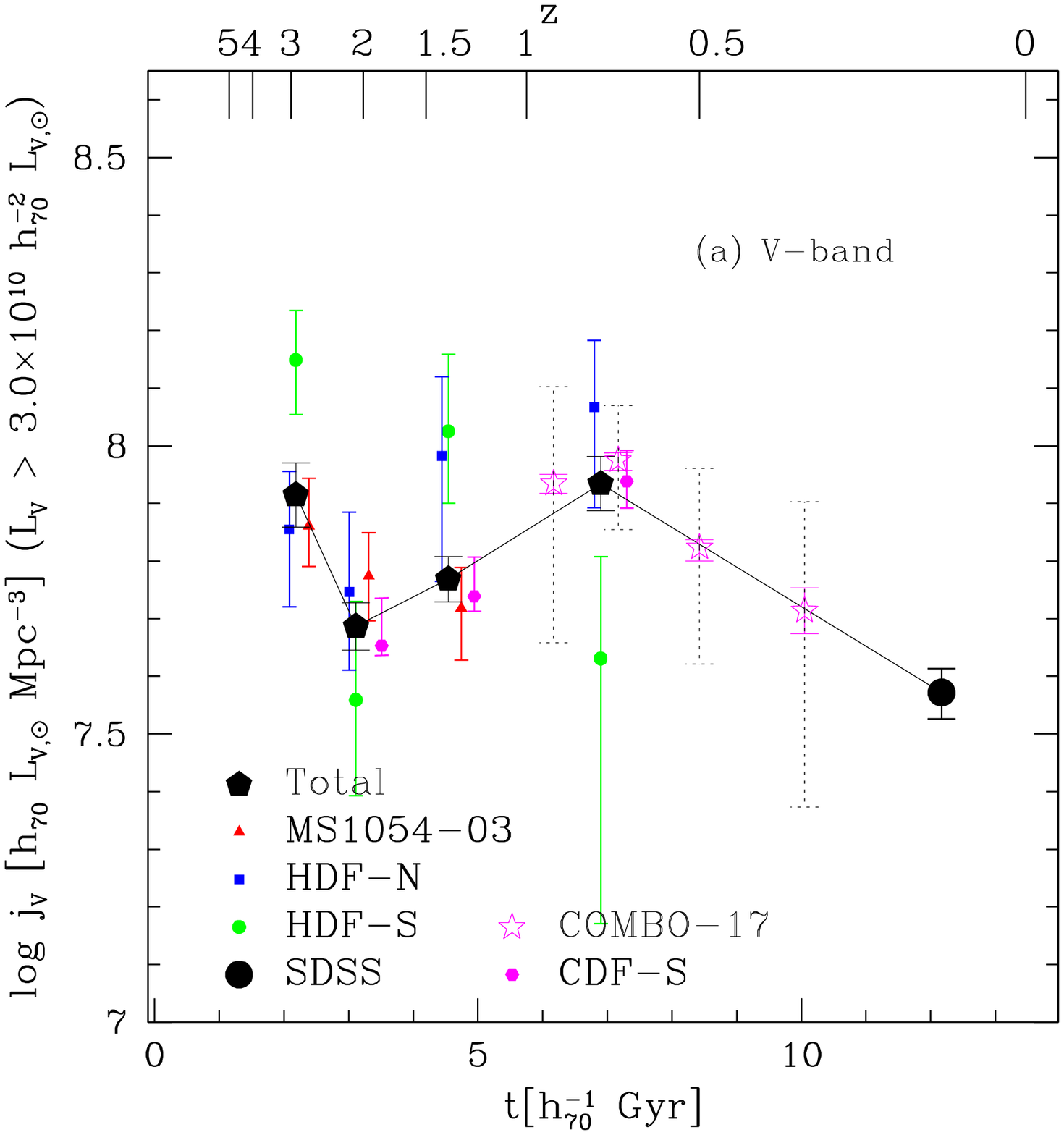}
\caption {The evolution of \jrestlam{V}, \jrestlam{B}, 
 and \jrestlam{U} with time in panels (a), (b), and (c) respectively
 for galaxies with \l{V}$>$\lthreshlam{V}.  The solid symbols indicate
 the values for the individual fields as labeled while the large solid
 symbol is the mean value.  The values are offset in time for viewing
 clarity.  The errorbars reflect the 68\% uncertanties in each field
 as characterized by a bootstrap simulation.  The dotted errorbars
 represent the explicit rms field-to-field variations among our four
 fields.  The open stars are the values for the COMBO-17 data and the
 open circle is the value for the SDSS.  The dotted errorbars in the
 COMBO-17 data reflect the field-to-field rms uncertainties among the
 3 COMBO-17 fields.  Note the large field-to-field variation even
 between fields as large as 0.25 square degrees.  Note also that the
 MS1054-03 field was excluded from the lowest redshift bin because of
 the presence of a rich cluster and that we excluded the CDF-S from
 the highest redshift bin due to the depth of the NIR data in that
 field, which which made us highly incomplete at
 \l{V}$>$\lthreshlam{V} and $z>2.41$.}
\label{lumdens_fig}
\end{figure}

 R01 were the first to estimate the evolution of the rest-frame
 optical luminosity function and luminosity density to $z\sim 3$,
 based on the number of rest-frame optically luminous galaxies in the
 HDF-S.  In Figure~\ref{lumdens_fig} we show \jrestlam{V},
 \jrestlam{B}, and \jrestlam{U} as a function of cosmic epoch for our
 four deep fields and for the complementary $z<1$ surveys.  These
 estimates are given in Table~\ref{tablum_UV} and
 Table~\ref{tablum_opt} .  As in R03, there is little evolution of
 \jrestlam{V} at $z<3$\footnote{The downward kink in \jrest\ at
 $z\sim2$ is likely due to systematic effects in the photometric
 redshifts, which tend to preferentially depopulate the $1.6<z<2$
 region.}.  Figure~\ref{lumdens_fig} also shows the large
 field-to-field variations present in our four deep fields.  Large
 variations in \jrestlam{V} among the three COMBO-17 fields are also
 seen, as indicated by the dotted error bars in
 Figure~\ref{lumdens_fig}.  Still, it is encouraging that the mean
 evolution in \jrestlam{V} is smoother than in the individual fields,
 indicating that we are getting closer to a measure of the true volume
 average value.

 The power of the COMBO-17 data is demonstrated here.  With its
 inclusion it is apparent that there is a slight evolution toward
 brighter \jrestlam{V} out to $z\sim 1$, as found by many authors
 (e.g. Lilly et al. 1996; Wolf et al. 2003).  Moving to bluer
 wavelengths the evolution from $z\sim 3$ to $z\sim0$ becomes more
 apparent, even within our own data.  This already foreshadows the
 trends with color that we discuss in the following section.

 The field-to-field rms in \jrestlam{V} ranges from $20-75\%$ in our 4
 redshift bins.  From this it is clear that data must be combined over
 large areas to robustly measure the evolution, especially for bright
 galaxies, which may be strongly clustered (e.g. Giavalisco \&
 Dickinson 2001; Daddi et al. 2003).  To mitigate field-to-field
 variations the most it is desirable for the individual fields to be
 larger than the typical clustering scale of objects in the field or
 to be spread over many independent lines of sight \citep{Somerville04}.

\begin{figure}
\ifemulate 
	\epsscale{1.2}
\else
	\epsscale{1.0}
\fi
\plotone{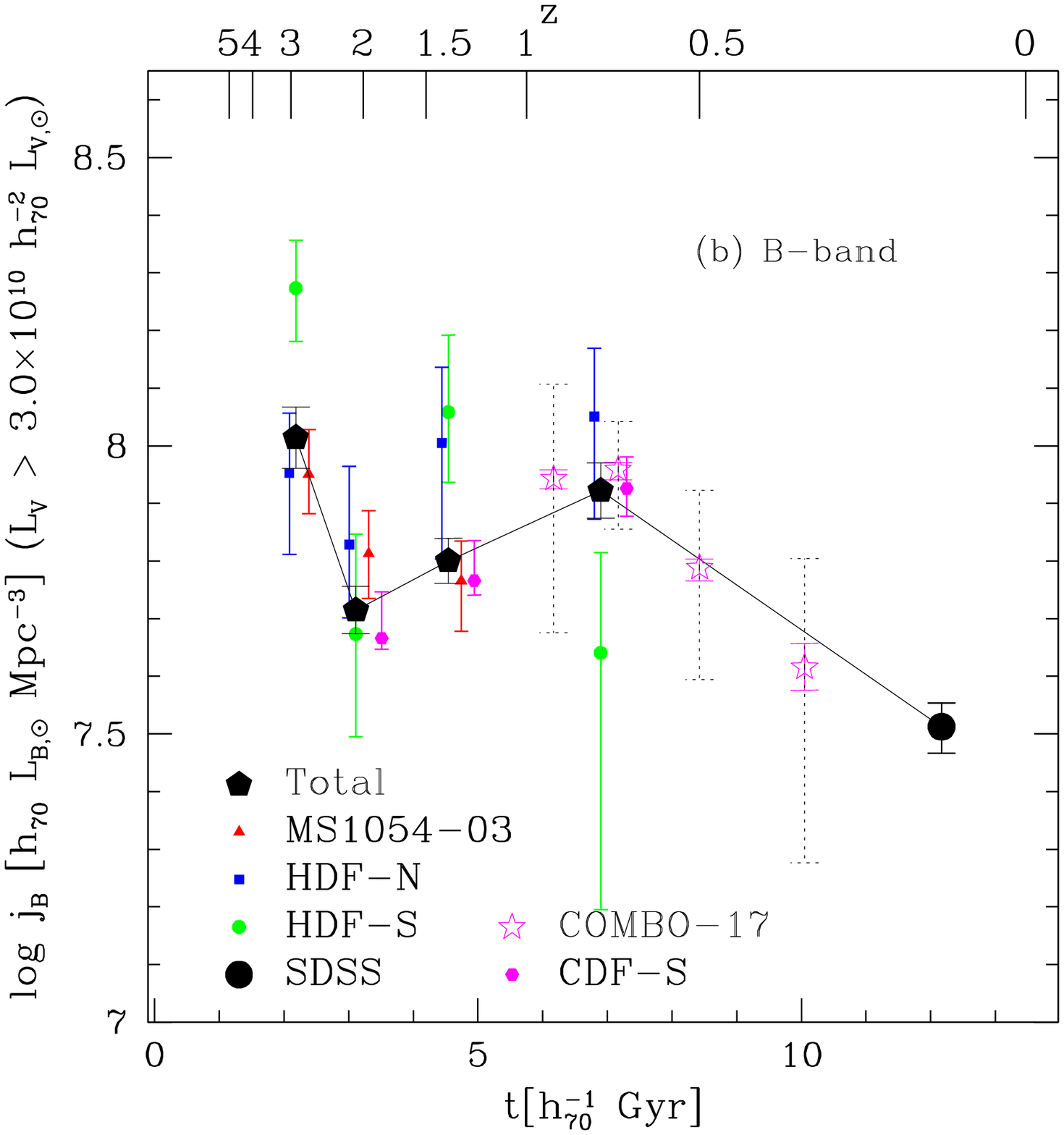}
\end{figure}

\begin{figure}
\ifemulate 
	\epsscale{1.2}
\else
	\epsscale{1.0}
\fi
\plotone{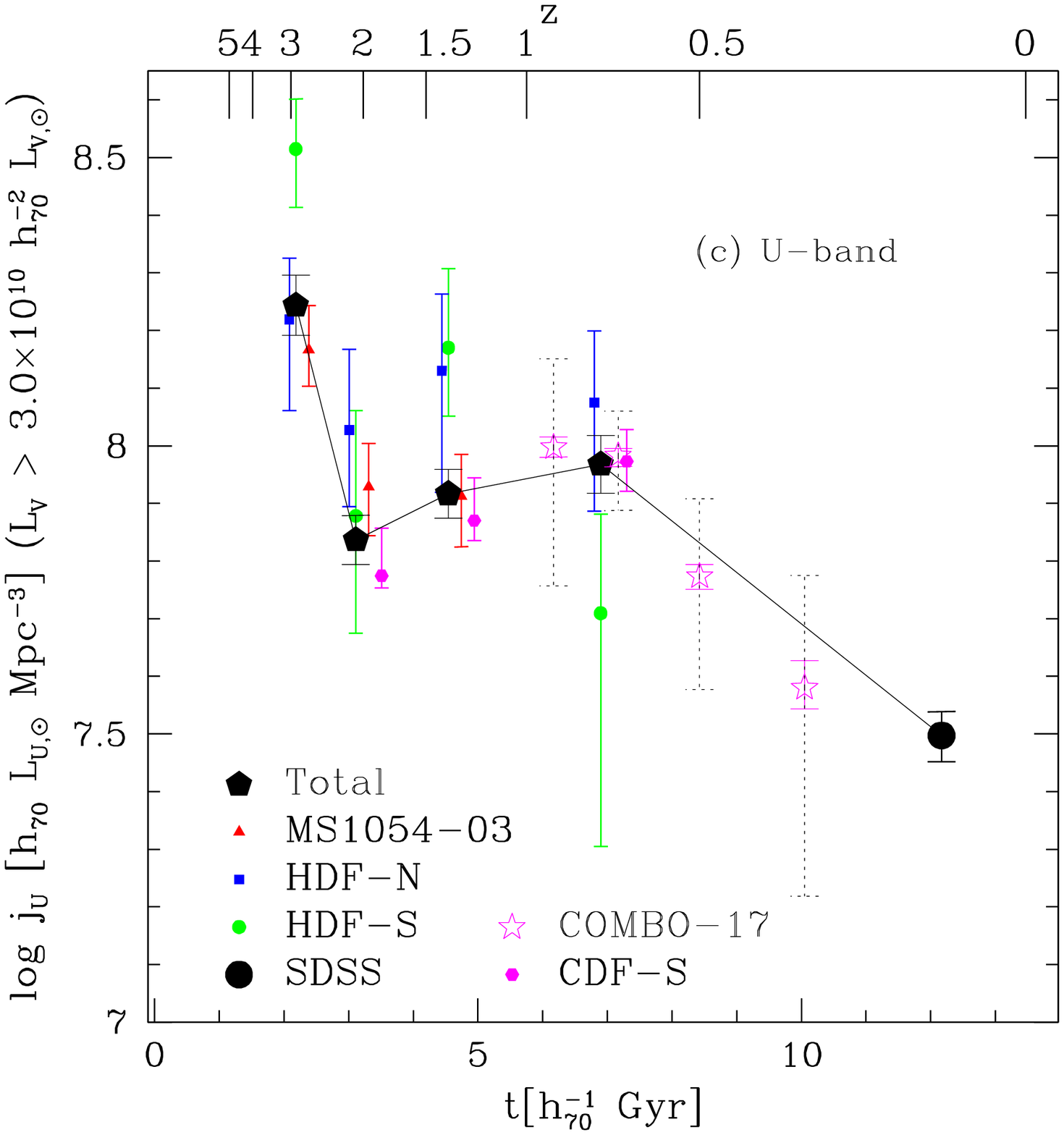}
\end{figure}

\subsection{The Volume Averaged Color and SED}
\label{SED_sec}

\begin{figure}
\ifemulate 
	\epsscale{1.2}
\else
	\epsscale{1.0}
\fi
\plotone{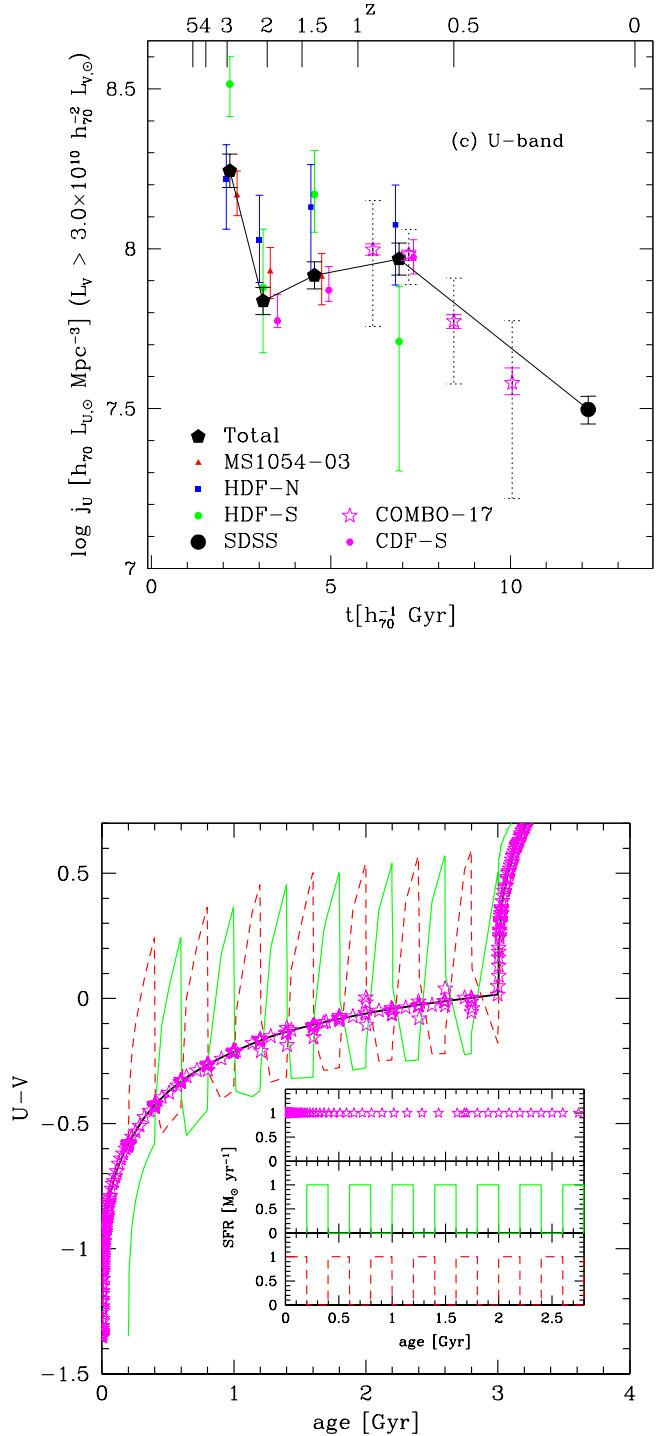}
\caption {A stylized example of how the volume averaged color of 
bursty SFHs reflects the sum of the individual SFHs.  In the inset
plot we show in the two lower panels two periodic SFHs with an
identical duty cycle and peak SFR but which are perfectly out of
phase.  In the top panel we show the sum of their SFHs, which is equal
to a constant.  The main plot shows the evolution of the \uv\ colors
associated with the individual SFHs and the evolution of a constant
SFR model which is simply the sum of the individual SFHs (thick black
line).  We also show the \uv\ color which is derived from adding up the
luminosities of each SFH in each band and using them to derive a total
color as described in the text (stars).  This luminosity weighted
color is exactly the same as the color of a constant star formation
model which is the sum of the individual SFHs.}
\label{burstcol_fig}
\end{figure}

 We now proceed to measure the galaxy-averaged rest-frame colors at a
 given redshift/epoch.  As discussed in R03 this average reflects a
 relatively smooth SFH for the ensemble of galaxies, even if
 individual SFHs are bursty.  Therefore, a conversion of rest-frame
 color into a mean \mlstar\ will be more robust when modeling with a
 smooth SFH.  In Figure~\ref{burstcol_fig} we illustrate this using an
 example which shows that the volume averaged color of a set of bursts
 behaves just like a population with the sum of their SFHs.

\begin{figure}
\ifemulate 
	\epsscale{1.2}
\else
	\epsscale{1.0}
\fi
\plotone{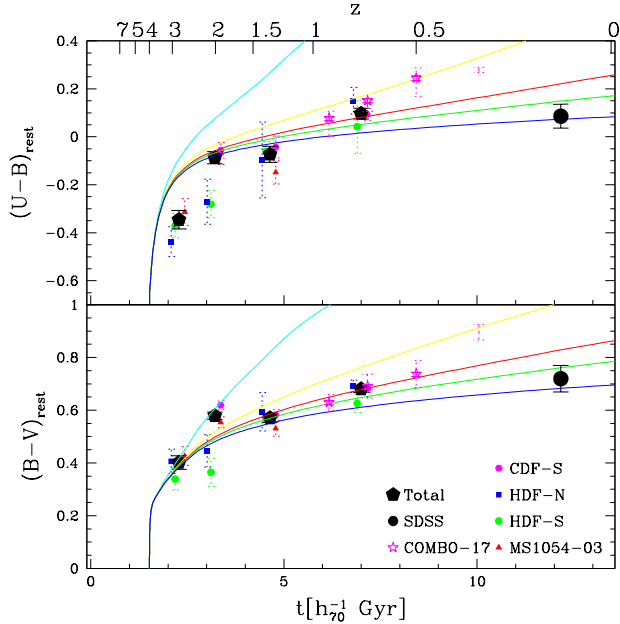}
\caption {The redshift evolution of the volume averaged \ubr\ and \bvr.  
The symbols are identical to those in Figure~\ref{lumdens_fig} with
the addition that the \ubr\ colors have also been corrected for
probable emission line contamination as in R03.  For clarity we have
also plotted the errorbars of the individual fields as dotted lines.
The colors of the lowest redshift COMBO-17 point are artificially red
due to the use of small central apertures to measure the color.  For
this reason we de-emphasize this point by only plotting a dotted
errorbar.  The integrated colors evolve redward with decreasing
redshift with a much smoother progression than seen in the evolution
of \jrestlam{V} in Figure~\ref{lumdens_fig}.  The solid tracks
indicate a set of solar metallicity exponentially declining SFHs with
$z_{form}=4$, $E(B-V)=0.25$ (using a Calzetti et al. 2000 attenuation
law), and with timescales of 1, 3, 6, 10, and 100 Gyr moving from top
to bottom.  Although these are merely example SFHs they highlight the
inability of simple models with an exponential SFH, dust content, and
metallicity to fit the time evolution of the global colors.}
\label{colz_fig}
\end{figure}

 In Figure~\ref{colz_fig} we present the redshift evolution of the
 luminosity weighted average color of all galaxies with
 \l{V}~$>3\times 10^{10}$\lumsol, computed using Equation~\ref{coleq}.
 The colors become continuously redder from $\sim 3$ to $z\sim 0$.

 Contrary to the conclusions of R03, with our multiple fields it is
 apparent that the colors are only slightly less susceptible to
 field-to-field variations than the luminosities, with the rms
 variations per redshift bin ranging from 5-50\% for \uvr\ and 20-75\%
 for the \jrestlam{V}.  This indicates that the mix of SED types also
 varies greatly from field to field, implying that large area surveys
 are even necessary to study the mean stellar populations of luminous
 galaxies in the Universe.

\begin{figure}
\ifemulate 
	\epsscale{1.2}
\else
	\epsscale{1.0}
\fi
\plotone{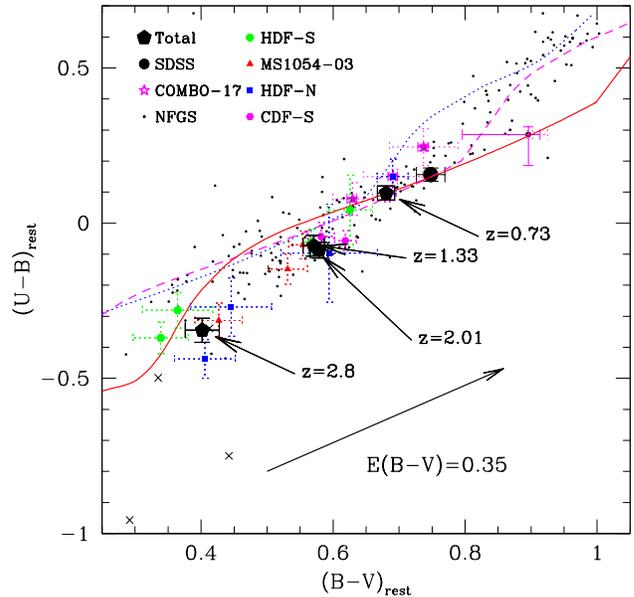}
\caption {
 The \ubr\ vs. \bvr\ at $z=$0.73, 1.33, 2.01, and 2.8 of all the stars
 in galaxies with \l{V}$>3 \times 10^{10}$\lumsol.  The large symbols
 are identical to those in Figure~\ref{colz_fig}.  For clarity we have
 plotted the errorbars of the individual fields as dotted lines and we
 do not plot the field-to-field errorbars for the COMBO-17 data.  The
 small solid points are the colors of nearby galaxies from the NFGS
 \citep{Jansen00a}, which have been corrected for emission lines.  The
 small crosses are the NFGS galaxies which harbor Active Galactic
 Nuclei.  The thin tracks correspond to an exponentially declining SFH
 with a timescale of 6~Gyr.  The tracks were created using a
 \citet{Sal55} IMF and the BC03 models.  The dotted track has no
 extinction, the dashed track has been reddened by $E(B-V)=0.15$, and
 the thin solid track has been reddened by $E(B-V)=0.35$, using the
 \citet{Calz00} extinction law.  The thick black arrow indicates the
 reddening vector applied to the solid model track.  The emission line
 corrected data lie very close the track defined by observations of
 local galaxies but are systematically bluer than all of the smooth
 models.}
\label{colcol_fig}
\end{figure}

 In Figure~\ref{colcol_fig} we show our average color estimates and
 those from COMBO-17 and SDSS in color-color space.  It is clear that
 all of the average estimates lie on a narrow locus in this space.  To
 interpret this trend we refer to \citet{Lar78} who demonstrated that
 individual galaxies with normal morphological types lie on a narrow
 locus in the \ubr\ vs. \bvr\ diagram and that galaxies with large
 bursts of star formation scattered away from the relation,
 preferentially to blue \ubr\ colors.  To ascertain the amount of
 ``burstiness'' in our average colors we therefore compare them to
 those of individual local galaxies from the NFGS (Jansen et
 al. 2000a) in Figure~\ref{colcol_fig}.  In general our colors and
 those of COMBO-17 are similar to the colors of local galaxies,
 implying a relatively smooth ensemble SFH.  Our data in the highest
 redshift bin, however, have a slight systematic offset to bluer \ubr\
 colors at a given \bvr\ color with respect to the local sample.  We
 explored whether part of this offset could result from systematic
 errors in the photometric redshifts.  At $z>2$ it appears that \zp\
 slightly overestimates the redshift on average, by about 0.2.  These
 uncertainties likely result from an imperfect template set or an
 incorrect amount of extinction.  Although the redshift solution is
 driven primarily by the rough position of the dominant break in the
 SED, small color mismatches in the template set can slightly change
 the redshift.  These systematic redshift errors can indeed produce
 \ubr\ colors which are bluer than the true values by up to 0.1
 magnitudes, although they can't entirely make up the differences with
 respect to the local data.  The \ubr\ color is especially sensitive
 to these effects, implying that more information than simply one
 color should be used to infer physical properties of the stellar
 population.  The remaining differences between the high redshift and
 local samples can be understood, at least qualitatively, by
 considering the effects of bursty SFHs on the mean color.  Although
 the deviation from a smooth SFH is reduced when averaging over more
 galaxies there is still some residual burstiness left.

\begin{figure}
\ifemulate 
	\epsscale{1.2}
\else
	\epsscale{1.0}
\fi
\plotone{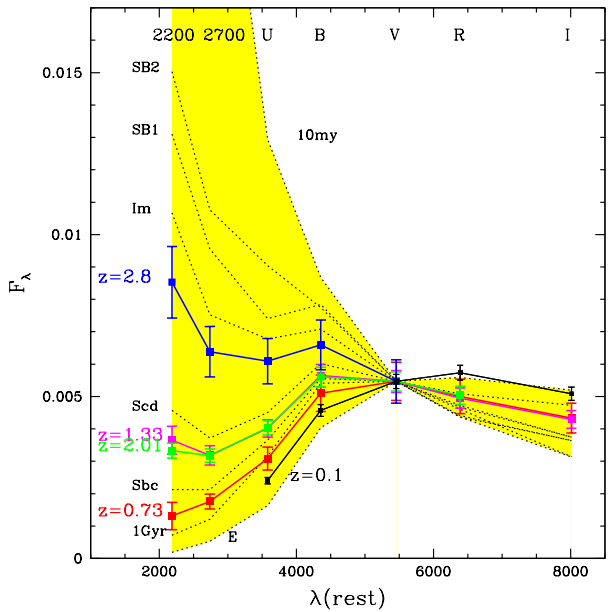}
\caption {The 2200\ang\ through \mi-band volume averaged SED of all the 
 galaxies with \l{V}~$>3\times 10^{10}$\lumsol.  The different colored
 lines correspond to the different redshifts.  The mean SEDs have been
 normalized to the same \mv-band rest-frame luminosity to highlight
 the evolution in the colors.  The dotted lines are the SEDs of a set
 of galaxy templates, as labeled on the left side.  The
 2200\ang\ through \mi-band volume averaged SED at all redshifts falls well
 within the range of colors occupied by ``normal'' nearby galaxies.
 Even at the highest redshift bin the mean SED looks more like that of
 a late type galaxy than of a starburst.  A break is also visible
 between the \u\ and \mb\ bands, indicating the presence of an evolved
 population - with an associated 4000\ang/Balmer break - that is
 dominating the light.  }
\label{sed_evolve_fig}
\end{figure}

 We can also use the rest-frame UV to optical volume averaged SED to
 further constrain the nature of the mean stellar population. In
 Figure~\ref{sed_evolve_fig} we show the galaxy-averaged SED from the
 rest-frame UV through the optical for our four redshift bins and
 compare it to the SEDs of local galaxy templates.  The volume
 averaged SED at all redshifts falls well within the locus of normal
 local galaxies.  Even in our highest redshift bin the mean SED is not
 as blue as a local starburst.  Also evident is the presence of a
 break in the mean SED between the \u\ and \mb\ bands.  That this break
 is present even in our highest redshift bin indicates that the
 rest-frame optical light at $z>1.6$ in rest-frame optically bright
 galaxies has significant contributions from evolved stellar
 populations.

\begin{figure}
\ifemulate 
	\epsscale{1.2}
\else
	\epsscale{1.0}
\fi
\plotone{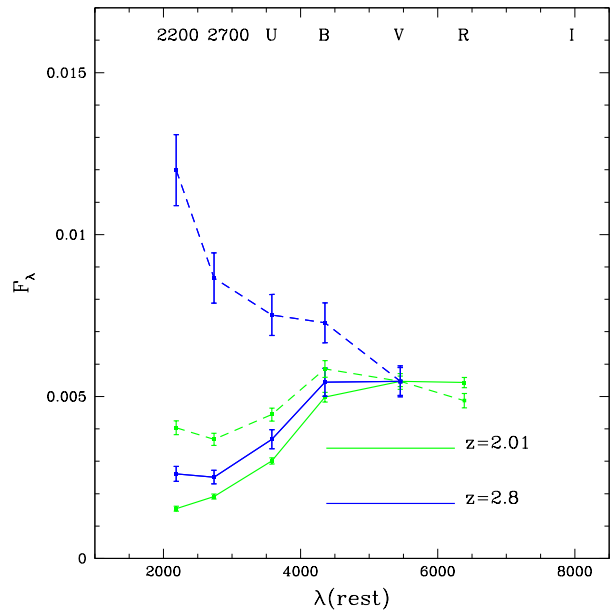}
\caption {Same as for Figure~\ref{sed_evolve_fig} but splitting the 
 two highest redshift bins split into DRGs (solid lines) and non-DRGs
 (dashed lines).  At both redshifts the DRGs are redder than the
 non-DRGs, both in the rest-frame optical - where breaks are easily
 visible - and in the rest-frame UV.  Because the DRGs still have
 light in the rest-frame UV the red colors are likely caused by a mix
 of old stars and dusty star formation.  }
\label{sed_drgcomp_fig}
\end{figure}

 In Figure~\ref{sed_drgcomp_fig} we also split our sample into DRGs
 and non-DRGs and display their volume averaged SEDs.  This is only
 done in the two highest redshift bins, covering the redshift interval
 targeted, by design, by the DRG selection criterion.  The DRG mean
 SED is uniformly redder than that of the non-DRGs.  This is not
 surprising at rest-frame optical wavelengths since these galaxies
 were in fact selected to have red rest-frame optical (observed NIR)
 colors.  The origin of the redder optical colors is partially
 elucidated by the detailed shape of the mean SED.  It is apparent
 that the DRGs have stronger rest-frame optical breaks than the
 non-DRGs, indicative that evolved stellar populations are more
 prevalent in DRGs than in non-DRGs, to the same rest-frame optical
 luminosity limit.  Similar conclusions were reached by
 \citet{Forster04} and L03 based on the results of SED fits to
 individual galaxies.  DRGs, selected by their red rest-frame optical
 colors also have red UV SEDs.

 In summary, it appears in all cases that the volume averaged SED of
 all rest-frame optically luminous galaxies has a strong contribution
 from evolved stellar populations and evolves to redder colors toward
 lower redshifts.  

\subsection{Modeling the Volume Averaged Stellar Population}
\label{modfit_sec}

 In order to derive global \mlstar\ and subsequently \rhostar\ values
 from our mean SEDs it is necessary to use models to interpret them.
 In R03 we derived \mlstarlam{V} estimates by using exponentially
 declining models applied to the global \uvr\ color in which the SFR
 timescale, dust content, and metallicity were constant as a function
 of redshift \footnote{In Figure 3 of R03 the tracks were mistakenly
 plotted with an extinction that varied with redshift.  These tracks
 were correctly plotted with a constant $E(B-V)$ in all other plots
 but the mistake in Figure 3 influenced our choice of best fit
 $E(B-V)$ and therefore affected our absolute determinations of
 \mlstar\ and \rhostar\ by about 0.1-0.2 dex.  The main result of R03,
 the relative change in \mlstar\ and \rhostar\ with redshift, however,
 was \textit{not} affected since all values were computed with the
 same model.}.  With better data we attempt this same exercise again
 and find that the same simple models cannot accurately describe our
 higher precision data.  We show five example models in
 Figure~\ref{colz_fig}.  These models have a \citet{Sal55} IMF, solar
 metallicity, $E(B-V)=0.25$, $z_{form}=4$ and exponentially declining
 SFHs with timescales of 1, 3, 6, 10, and 100 Gyr.  Although these
 models represent only one possible set of parameters we find it is
 generally impossible to simultaneously fit the \ubr\ and \bvr\ colors,
 in the sense that the models have consistently too red \ubr\ colors
 for a given \bvr\ color for our data but with the opposite trend
 apparent for the COMBO-17 data.  This same disagreement is seen in
 Figure~\ref{colcol_fig} where we plot our data, along with that of
 COMBO-17 and SDSS, in the \ubr\ vs \bvr\ color plane.  The emission
 line corrections we applied to the colors did move the data in the
 direction of better agreement with the models but these corrections
 are already quite extreme at high redshift (0.04 magnitudes in \ubr)
 and larger shifts would be difficult to accommodate given the range
 in corrections inferred from the NFGS\footnote{Of course, this is
 under the assumption that the EWs of emission lines in the NFGS are
 the same as those of high redshift galaxies at similar rest-frame
 colors.  If, e.g. high redshift galaxies have higher equivalent
 widths then it would imply that we are undercorrecting for emission
 lines.}.  To determine the robustness of the disagreement between the
 models and data we have explored different combinations of
 metallicity and extinction values and have found that no single set
 can reproduce these two colors at all redshifts.  As we have
 mentioned already, however, the \ubr\ color in particular is
 susceptible to redshift uncertainties and uncertainties in the
 emission line corrections.

 To mitigate the influence of one color on the global fit we choose to
 model the full UV-optical volume averaged SED to constrain the set of
 population parameters.  As we have pointed out earlier, averaging
 over the whole galaxy population also averages over the SFHs of the
 individual galaxies, making it more appropriate to apply smoothly
 varying SFHs when performing the model fits.  This point of view is
 generally supported by the general agreement of our mean SEDs with
 those of ``normal'' local galaxies, which have been shown to be
 consistent with extended, relatively smooth SFHs (e.g. Kennicutt,
 Tamblyn, \& Congdon 1994).  With our choice of simple models there
 are four possible free parameters that we consider: age, metallicity,
 dust, and SFH.  We use the BC03 models and an updated version of the
 \textit{Hyperz} code \citep{Bolz00} to fit our data (Bolzonella
 private communication).  We assume a Salpeter IMF with lower and
 upper mass cutoffs of 0.1 and 100 \msol\ respectively.  Fitting with a
 different IMF, e.g. a Chabrier, would result in a scaling of the
 \mlstar\ values by a constant factor at all redshifts (assuming that
 the IMF is universal).  Thus the relative trends would be preserved.
 In all cases we use the \citet{Calz00} attenuation law.  We find that
 there exists some combination of parameters that can fit the
 rest-frame UV-optical volume averaged SEDs as long as we don't
 require that the model parameters are the same at all redshifts.

 When fitting broad band photometry, especially only out to the
 rest-frame optical, there are very large degeneracies between age,
 SFH, metallicity, and dust content, e.g. P01, \citet{Shap01}.  For
 example, in some cases high redshift SEDs can be fit reasonably well
 with both a constant SFH (hereafter CSF), moderate ages, and high
 extinctions and SFRs, or with an exponentially declining model with
 lower extinction and SFRs.  Despite these degeneracies, however,
 \mlstar\ are better constrained, since different combinations of
 stellar population parameters can give similar derived mass-to-light
 ratios \citep{Forster04, Dokkum04}.  We will therefore use our fits
 to measure \mlstar\ but will not discuss the constraints on the
 stellar population parameters in detail.  There is tentative evidence
 from NIR spectroscopy of bright galaxies at $z<3$ that metallicities
 are typically around solar \citep{Dokkum04, Shap04, Mello04} and we
 therefore limit our fits to solar metallicity.

 To span the range of possible models we perform fits with either a
 CSF, Simple Stellar Population (SSP), or exponentially declining
 model with a timescale of 300 Myr (\taumod{300Myr}).  In all cases we
 restrict the age of the population in each redshift bin to be younger
 than the age of the Universe for the middle redshift in that bin.  We
 parameterize the amount of extinction by the magnitudes of
 attenuation in the \mv-band $A_V$.  For the CSF and \taumod{300Myr}\ models we
 allow the extinction to vary between $0<A_V<3.0$ and for the SSP
 models between $0<A_V<0.5$.  This latter choice is to prevent the
 fitting from choosing highly extincted very young models with no
 ongoing star formation, which would result in unphysically large
 intrinsic luminosities.  The upper limit of 3.0 magnitudes of
 extinction does not affect our conclusions.  For each redshift bin we
 choose the best-fit SFH, age, and extinction combination and use the
 \mlstarlam{V} of that model, which includes the stellar mass loss
 from evolved stars.  We then multiply that \mlstarlam{V} by
 \jrestlam{V} to derive \rhostar.

 To derive uncertainties on \mlstarlam{V} and \rhostar\ we fit the
 ensemble of mean SEDs produced by our bootstrap simulation described in
 \S~\ref{ldens_meas_sec}.  Each bootstrapped SED was computed from
 the same set of galaxies, which is necessary because each galaxy
 contributes luminosity in every passband, i.e. the \jrest\ values are
 correlated.  We then use the distribution of \mlstarlam{V} and
 \rhostar\ values from the fit to the ensemble of SEDs and determine
 the 68\% confidence limits.

\begin{figure*}
\epsscale{0.9}
\plotone{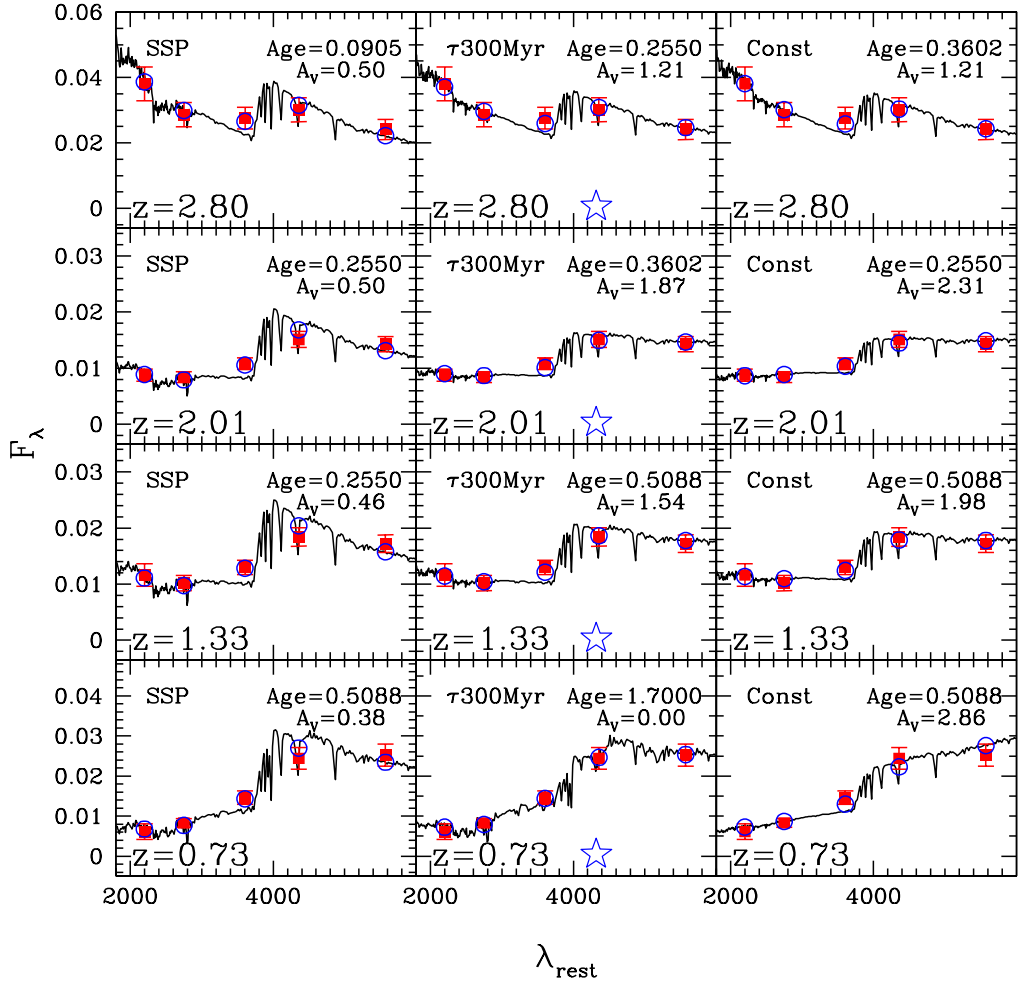}
\caption {The best fit models to the volume averaged SED.  
 Each row corresponds to one of the redshift bins and each column
 corresponds a different SFH.  The panel in each row with a star
 indicates the formal best fit SFH, although the other SFHs are
 generally allowed within the uncertainties.  The filled points show
 the measured values and the open circles show the best fit model
 fluxes.  As in Figure~\ref{sed_evolve_fig} the rest-frame optical
 break of both the data and best-fit model is apparent at all
 redshifts and increases toward lower redshifts.  }

\label{sed_fit_fig}
\end{figure*}

 The best fit SFH for each volume averaged SED is presented in
 Figure~\ref{sed_fit_fig}.  We also indicate the formal best fit SFH,
 which is the \taumod{300Myr}\ model for the three lowest redshift bins and the
 SSP model for the highest redshift bin.  It is important to note,
 however, that the \taumod{300Myr}\ model is statistically allowed at all
 redshifts and that our \mlstar\ determinations are not very sensitive
 to the exact SFH.  As was visible in Figure~\ref{sed_evolve_fig} the
 rest-frame optical break is present at all redshifts, with the
 strength of the break decreasing toward higher redshifts.  This is
 also reflected in the model fits.

 We also fit the mean SED from the SDSS survey to constrain the
 evolution of \mlstar\ and \rhostar\ to low redshift.  The COMBO-17 data
 were very useful for characterizing the overall trends of color and
 for assessing the applicability of certain models for the color
 evolution.  Nonetheless, we do not model the COMBO-17 data in detail
 since the lack of deep NIR observations limits us to only a
 relatively small range in rest-frame wavelength at each redshift,
 with correspondingly poor constraints on the best-fit \mlstarlam{V}.
 The low redshift sample from the SDSS is crucial as our \l{V} cut
 complicates the use of literature values for \rhostar, which are
 always quoted as total values.  By fitting the mean SED of the local
 sample as determined with our \l{V} threshold we can consistently
 track the evolution over redshift.  The only differences in analyzing
 the low and high redshift samples is in the allowed stellar
 population parameters and the derivation of uncertainties.  Because
 luminous galaxies from the SDSS are almost entirely made up of
 evolved early type populations \citep{Blanton03a, Kauff03} we limited
 the attenuation to 1 magnitude in the \mv-band.  In addition it is
 difficult to obtain a realistic uncertainty estimate for the SDSS.
 The uncertainties in \jrest, and hence in \mlstar\ and \rhostar\ are
 dominated by systematic uncertainties.  Therefore we give a 10\%
 error to all derived SDSS quantities.
 
\subsection{\mlstarlam{V} and \rhostar}
\label{massevolve_sec}
 
 In Figure~\ref{ml_fig} we show the evolution of the derived
 \mlstarlamav{V} for luminous galaxies as a function of time and
 redshift.  These are also given in Table~\ref{mltab}.  As expected
 the bluer colors of the volume averaged SEDs and the decrease in the
 break strength toward higher redshift implies a lower value of
 \mlstarlamav{V}.  \mlstarlamav{V} increases by a factor of $\sim 10.7$
 from $z=2.8$ to $z=0.1$.

 Using the \mlstarlamav{V} values and the \jrestlam{V} measurements we
 derive \rhostar\ for galaxies at high redshift and in the SDSS with
 \l{V}$>$\lthreshlam{V}.  Our absolute \rhostar\ measurements at
 $0<z<3.2$ are given in Table~\ref{masstababs}.  The mass density in
 luminous galaxies between decreases by a factor of $\approx 4.8$
 between $z=0.1$ and $z=2.8$ and $\sim 50\%$ of the stellar mass was
 in place by $z\gtrsim 1$.  This is in broad agreement with the
 results of R03, who measured a decline by a factor of $\sim 10$ in
 \rhostar\ between $z=0.1$ and $z=2.8$, although with significantly
 larger errorbars and with no accounting for cosmic variance.  The rms
 field-to-field variations in \rhostar\ among our four fields ranges
 from $40-61\%$.

 To compare our data to that from other authors we scale our data to
 ``total'' values.  We calculate an upwards correction factor of 2.1
 by measuring the total SDSS \rhostar\ value and comparing it to
 \rhostar\ for the SDSS galaxies with \l{V}$>$\lthreshlam{V}.  After
 correcting all our measurements to total we normalize our values to
 the $z=0$ \rhostar\ measurement of C01 and plot them along with those
 of other authors in Figure~\ref{massdens_fig}.  The relative values
 are given in Table~\ref{masstab}.  We also determine how the inferred
 evolution in the total \rhostar\ changes if we allow \lthreshlam{V}
 to evolve in the same way as our \mlstarlam{V} values.  The change
 this can induce is indicated by the vertical extent of the boxes in
 Figure~\ref{massdens_fig} and is reflected in the confidence interval
 in $\Delta\rho_\star$ given in Table~\ref{masstab}.  Our local total
 mass density measurement from SDSS is $\approx 25\%$ lower than that
 of C01.  Some of this discrepancy may be due to the slightly higher
 mean redshift of our data.  C01 estimate their stellar masses at
 $z=0$ whereas the mean redshift for the SDSS sample is 0.1.  Using
 the $z=0.1$ SFRD from \citet{Brinch04} we would predict that
 \rhostar\ would increase by $\approx 9\%$ between $z=0.1$ and 0,
 accounting for some of the discrepancy.  Another possible source of
 the discrepancy is the use by C01 of a fixed formation redshift of
 $z=20$ for their SED fitting derived \mstar\ estimates.  

 According to \citet{Bell03} the old age assumed for all galaxies
 could bias their estimates by $\approx 10\%$
 \citet{Bell03}\footnote{The \rhostar\ determination of \citet{Bell03}
 agrees excellently with that of C01 and also assumes a fixed old age
 (12 Gyr) when deriving \mstar\ of galaxies in their sample.}.  By
 refitting the SEDs of galaxies with SDSS and 2MASS data and letting
 the age vary \citet{Fontana04} estimated that forcing an old age can
 bias the masses from C01 upwards by $<20\%$.  Taking all these
 effects into account, the remaining difference between our SDSS point
 and that of C01 falls well within the range of expected systematic
 errors in our \rhostar\ estimates.

Compared to the C01 point, the
 mass density at $z=2.8$ is 5--8 times less than at $z=0$, where the
 range corresponds to our formal 68\% confidence limits for a fixed
 \lthreshlam{V}.  As discussed in \S\ref{syserr_subsec}, however, it
 may not be appropriate to compare the evolution of our \l{V} limited
 sample to that of the total estimates presented by different authors
 (see below).  If we instead allow \lthreshlam{V} to vary along with
 our \mlstarlam{V} values the mass density at $z=2.8$ is constrained
 to be 5.3--16.7 times less than at $z=0$.

 At higher redshifts \rhostar\ is estimated by other authors using SED
 fits of individual galaxies.  In all cases we have converted to a
 Salpeter IMF when necessary.  \citet{Brinch00} fit a set of $I$-band
 selected galaxies at $z<1$ from the Canada France Redshift Survey
 (CFRS).  They are only complete over the range $10.5 < {\rm log}
 {\cal M_\star} < 11.6$ and their value here includes a $20\%$
 correction to a total mass density recommended by \citet{Brinch00}.
 \citet{Drory04} estimate \mstar\ for $z<1.2$ galaxies with $K<19$
 over 0.28 deg$^2$ from the Munich Infrared Cluster Survey (MUNICS).
 They also adopt an SED fitting technique but use their best-fit
 models to derive \mlstarlam{K} and then extrapolate their SEDs from
 the observed to the rest-frame $K$-band to derive \l{K}.  Using the
 DEEP2 spectroscopic survey coupled with \mk-band imaging,
 \citet{Bundy06} use an SED fitting technique to estimate galaxy
 masses and then fit a Schechter function and extrapolate it to get
 their \rhostar\ estimates.  At high redshifts/faint magnitudes, they
 supplemented their spectroscopy with photometric redshifts but these
 have a minimal effect on the derived \rhostar\ values.  The errorbars
 for the \citet{Bundy06} points are from the variance in \rhostar\
 between their four fields.  \citet{Borch06} computed the stellar mass
 density from 25,000 galaxies in the COMBO-17 survey.  They fit their
 combination of 17 broad and medium optical bandpasses with SEDs to
 derive the stellar masses for each galaxy.  They then fit Schechter
 functions to their data in each redshift bin and integrate this to
 derive \rhostar.  \citet{Fontana04} use the K20 survey to measure the
 stellar mass function out to $z=2$ using stellar masses derived by
 SED fitting of individual galaxies.  They then integrate this mass
 function to obtain \rhostar.  In their $1.5<z<2$ bin their mass
 function is complete only down to the $\sim {\cal M^\star}$ value of
 their best-Schechter fit to the mass function and the extrapolation
 to lower \mstar\ is large and uncertain.  D03 and F03 derive
 \rhostar\ at $z\sim3$ from the HDF-S and HDF-N WFPC2 fields
 respectively, where the raw data in the HDF-S are in common with the
 FIRES data and where the HDF-N catalog is identical to what is used
 in this paper.  The techniques in both these papers are nearly
 identical, using SED fitting of individual galaxies to determine the
 mean \mlstar\ for all galaxies and then applying this to the
 luminosity density as determined by integrating a Schechter function
 fit to the rest-frame optical luminosities.  \citet{Drory05} estimate
 the mass density in the FORS Deep Field (FDF) using an \mi-band
 selection sample and in the GOODS-S field using a \mk-band selected
 sample drawn from the same raw data as used for this paper.  They
 derive mass densities by fitting the observed SEDs and then add up
 all the galaxies in their sample at each redshift.  Although they
 measure \rhostar\ out to $z\sim 5$ we only plot it out to $z\sim 2$
 for two reasons. First, their \mi-band selection in the FDF
 corresponds to a selection in the UV at $z<1.5$ and the authors are
 therefore more sensitive to unobscured star formation than to stellar
 mass.  They point out that their \mi-band data is deep enough to
 detect all but 10\% of the \mk-selected objects in the deep FIRES
 images of the HDF-S but it is exactly these missed objects which
 comprise a significant fraction of the mass density at high redshift,
 i.e. DRGs.  Second, although they do have \mk-band data in the
 GOODS-S field it is not deep enough to be mass complete for any but
 the most massive objects at $z>2$, but they add up all objects,
 regardless of whether they are complete for those masses.  It is
 therefore difficult to interpret their total values at $z>2$.
 \citet{Gla04} determine stellar masses by fitting the observed SEDs
 with stellar population synthesis models and then compute \rhostar\
 using a $V/V_{max}$ method.  The compute \rhostar\ down to different
 mass limits and note their decreasing incompleteness in mass with
 increasing redshift and decreasing mass.  After converting to a
 Salpeter IMF we plot their points down to their lowest mass limit,
 $10^{10.45}$\msol.  At the highest redshift bins they are incomplete
 and the resultant \rhostar\ values can only be considered as lower
 limits.

 In our $0<z<1.0$ redshift bin our data fall slightly lower than the
 other determinations.  This difference may be because the CDF-S is
 underdense at low redshifts.  We suspect this because \citet{Wolf03}
 note that the Extended CDF-S is underdense at $0.2<z<0.4$ and that
 this trend apparently is true at $z\lesssim 0.6$ (Papovich private
 communication).  Although there is a loose structure in the Extended
 CDF-S at $z=0.67$ and a cluster at $z=0.73$ \citet{Gilli03}, these
 should not dominate the counts over the much larger CDF-S field.  We
 don't include the MS1054 data in our $0<z<1$ measurement because of
 the massive cluster in that field.  However, the large field-to-field
 variation among our 4 fields at $z<1$ makes our data consistent with
 those from other surveys.  At $z>1$ our \rhostar\ estimates agree
 very well with those of the other authors but with smaller formal
 errors corresponding to the lower uncertainties afforded by the
 multiple fields.  The only notable discrepancy is with the HDF-N
 determinations of D03; we address a possible cause for this in the
 following subsection.

 We also show the curves that correspond to the integral of the
 parametric fit to the SFR($z$) curve from (Cole et al. 2001; C01),
 both with and without a substantial extinction correction\footnote{As
 in R03 both curves have been corrected for the mass loss from evolved
 stars, which asymptotically approaches 30\% at 13 Gyr for a Salpeter
 IMF.  Changing the IMF to that of \citet{Chabrier03} would scale both
 the \rhostar\ and SFR($z$) measurements down by a factor of
 $1.4-1.8$.  In addition, the mass loss for a Chabrier IMF approaches
 50\% at 13 Gyr and results in a slightly different shape of the
 integral of the SFR($z$) curve.}.  As pointed out by, e.g.  D03, R03,
 F03, and \citet{Fontana04}, an extinction correction to the
 UV-derived star formation rates is required to match the \rhostar\
 measurements at all redshifts, although the nature of this correction
 is highly uncertain.  The \rhostar\ evolution predicted from the UV
 selected extinction corrected SFR is in broad agreement with the
 direct \rhostar\ measurements, notably at $z\sim 3$.  Nonetheless,
 the measurements, including those from the literature are
 systematically below the extinction corrected curve by about 0.2-0.4
 dex at $z<2$, with the exception of the \citet{Drory04} data.  This
 slight offset could be due to an erroneously high extinction
 correction to the SFR estimates, but these can't be changed too much
 without coming into disagreement with the local \rhostar\
 determinations.  Additionally, some of the disagreement may come from
 systematic underestimates of \rhostar, although it is not immediately
 clear what effects would plague all surveys, which use different
 techniques to estimate the mass densities.  Finally, it may be that
 the galaxies which enter into the SFR($z$) determinations are too
 faint in the rest-frame optical to enter the NIR selected samples.
 The star formation rate density (SFRD) measurements rely on an
 extrapolation to the faint end of the rest-frame UV luminosity
 function.  Estimates of the faint end slope have a large range, with
 the original determination from the HDF-N by \citet{Stei99} being
 $\approx-1.6$ but with a later measurement from the FORS Deep Field
 giving a value of $\approx-1.1$ \citep{Gabasch04}.  Also, as we
 describe in \S\ref{syserr_subsec} the \rhostar\ measurements can have
 systematic errors at the factor of $\sim 2$ level which could also
 account for the difference.  Obviously resolving the discrepancy
 between these two curves rests as much on an accurate determination
 of the SFR evolution as on that of \rhostar.

\subsubsection{The Stellar Mass Budget}
\label{massbudget_sec}

\begin{figure}
\ifemulate 
	\epsscale{1.2}
\else
	\epsscale{1.0}
\fi
\plotone{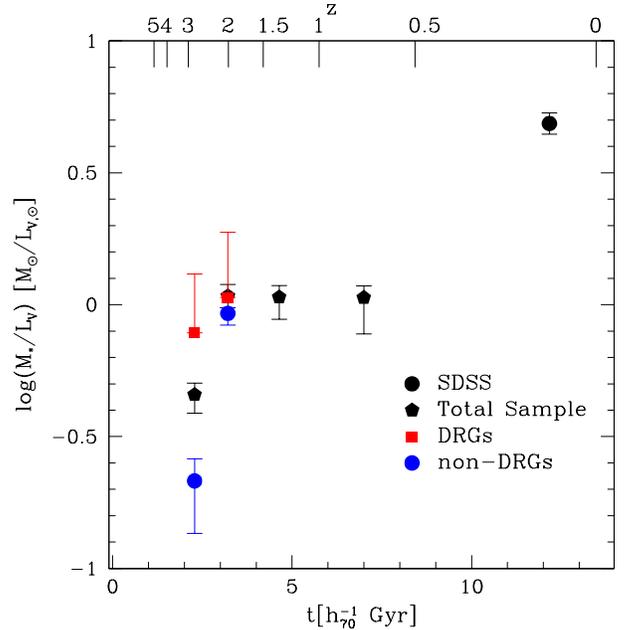}
\caption {The evolution of \mlstarlam{V} for luminous galaxies with 
 time and redshift, as derived by model fits to the volume averaged
 SEDS.  The black pentagons are for the entire \l{V}$>$\lthreshlam{V}
 sample and the blue circles and red squares are for non-DRGs and DRGs
 respectively.  The open circle represents the local determination
 from the SDSS. }
\label{ml_fig}
\end{figure}

\begin{figure}
\ifemulate 
	\epsscale{1.2}
\else
	\epsscale{0.75}
\fi
\plotone{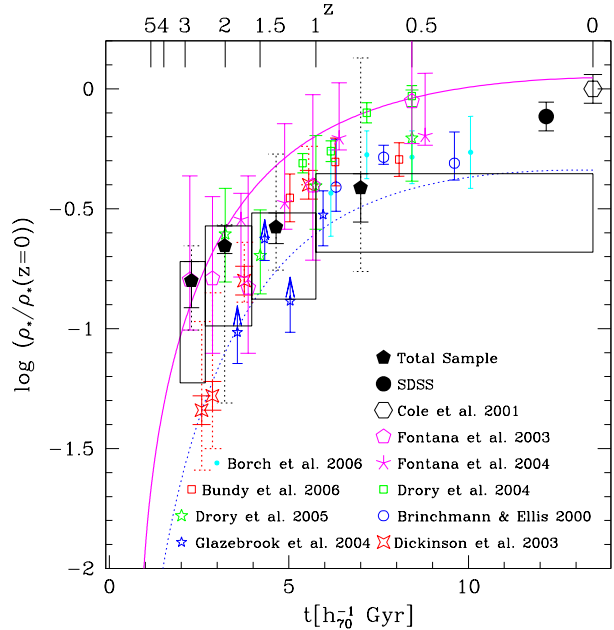}
\caption {The relative evolution of \rhostar\ with time and redshift. 
 The symbols are as indicated in the plot.  All quantities have been
 normalized to the local stellar mass density measured from
 \citet{Cole01}.  As described in the text, most other authors have
 quoted ``total'' values determined by extrapolating fits to the
 luminous/massive galaxy population.  Our measurements, on the other
 hand, are explicitly restricted to the luminous galaxy population,
 and are complete at \l{V}$>$\lthreshlam{V}.  The dotted errorbars on
 the solid pentagons indicate the minimum and maximum \rhostar\ values
 among our fields and given an indication of the field-to-field
 variation.  The horizontal extent of the solid boxes indicates the
 redshift range over which of our points are derived.  The vertical
 extent of the solid boxes indicates the range in \rhostar\ values
 that are obtained by allowing the rest-frame luminosity threshold to
 evolve to fainter luminosities at lower redshifts in accordance with
 our derived evolution in \mlstarlam{V}.  The mass density averaged
 over our four fields agrees well with most other surveys.  The one
 notable exception is the HDF-N (D03) which has a substantially lower
 value than either the HDF-S or the mean value computed here.  The two
 curves are the integrals of the \citet{Cole01} parametric fit to the
 SFR($z$) data.  The solid magenta line is the fit to the extinction
 corrected data and the dotted blue line line is for the data without
 an extinction correction.  As has been noticed by many authors an
 extinction correction to the SFR($z$) measurements is necessary to
 reproduce the \rhostar\ measurements at almost all redshifts.  There
 is a slight systematic difference between the \rhostar\ estimates and
 the integral of the SFR($z$) curve.  }
\label{massdens_fig}
\end{figure}

 With the larger number of galaxies afforded by our large area we can
 now split the sample into DRGs and non-DRGs to determine their
 relative contributions to the stellar mass budget.  Using the mean
 SEDs for these two subsets we calculate \mlstarlam{V}, which is shown
 in Figure~\ref{ml_fig}.  The mean \mlstarlam{V} for DRGs is a factor
 of 1.1 and 3.6 higher than for the non-DRGs at $z=2.01$ and 2.8
 respectively.  The higher \mlstar\ values for DRGs are in qualitative
 agreement with the results from SED fitting of individual galaxies
 \citep{Forster04, Dokkum04, Labbe05}.

 We show the corresponding mass densities as the blue and red filled
 pentagons in Figure~\ref{masssim_fig}.  The DRGs contribute 30\% and
 64\% of the stellar mass density at $z\sim2$ and 2.8 respectively,
 comparable to that from UV selected samples.  As shown in
 \citet{Franx03} and more recently by \citet{Reddy05} and
 \citep{Dokkum06} DRGs are almost entirely absent in rest-frame UV
 selected samples like the LBG or BM/BX samples.  Yet, they make up a
 comparable fraction of the mass budget down to similar rest-frame
 optical limits, showing that rest-frame UV selection misses
 significant amounts, if not most, of the stellar mass at $1.6<z<3.2$.
 Therefore NIR selection is crucial to obtaining a comprehensive and
 unbiased view of the high redshift Universe.  The important
 contribution of DRGs agrees with new results from \citet{Dokkum06}
 who show that they are the most numerous constituent of the
 population of galaxies with \mstar$>10^{11}$\msol\ and $2<z<3$.  It
 may be that the relatively unobscured star-forming UV-selected
 galaxies are over-represented in our luminosity selected sample due
 to their lower mass-to-light ratios.  If this is true then the
 contribution of DRGs in a mass-selected sample should therefore be
 higher.

 Interestingly the mass density of non-DRGs in our highest redshift
 bin is very similar to the HDF-N, which is $\sim 2$ times underdense
 compared to the total mass density from this work and from F03 and
 R03.  This may indicate that much of the field-to-field variation at
 high redshift originates in the population of red massive galaxies.
 This was already indicated by the higher clustering amplitude of red
 galaxies seen by \citet{Daddi03} in the HDF-S FIRES field and is also
 found by \citet{Dokkum06} who show that the CDF-S is underdense in
 massive (\mstar$>10^{11}$\msol) galaxies at $2<z<3$.  Clustering
 measurements of galaxies at high redshift as a function of color and
 over a much larger area will directly address this issue (Quadri et
 al. in preparation).

\subsection{Possible Systematic Errors and Biases}
\label{syserr_subsec}

 There are various systematic errors which can affect our conclusions.
 First is the limitation that our observations only probe out to the
 rest-frame \mv-band in our highest redshift bin.  While stellar
 mass-to-light ratios are much less variable in the rest-frame optical
 than the rest-frame UV they can still vary by an order of magnitude
 or more at a given \l{V}, depending on e.g. SFH, age, and extinction.
 Also, the extinction in very dusty starbursts can still be quite high
 in the rest-frame optical, implying that the light (and hence mass)
 from highly extincted stars will be missed.  The work of
 \citet{Labbe05} and \citet{Shap05} constrain the possible errors by
 deriving \mstar\ for objects using rest-frame NIR data
 obtained with the IRAC instrument on the Spitzer Space Telescope.
 For the most part \mstar\ derived from these fits agree with those
 derived from fitting only out to the rest-frame optical, although the
 \mstar\ uncertainties are reduced with the longer wavelength baseline.
 This implies that most of the stars that are visible in the
 rest-frame NIR are also visible in the optical, even though they may
 have a higher extinction.  If significant stellar mass exists in
 environments which are obscured at optical and NIR wavelengths then
 our results would be biased.  The amount of such heavily obscured
 stellar mass at $z<1$ is small, but increases rapidly out to higher
 redshift \citep{Lefloc05}.  If the fraction of extremely obscured
 stellar mass in rest-frame optically luminous objects at $z>1$
 becomes significant then the decline in the true mass density will be
 shallower than what we observe.

 Another systematic effect may stem from our relatively bright
 magnitude limits.  Even in the HDF-S, which has the deepest \Ks-band
 data in existence, we are limited to probing high luminosity galaxies
 and therefore may not be measuring a representative sample of the
 full galaxy population.  We estimate the amount of light that we miss
 with our \l{V} limit by using the rest-frame $B$-band luminosity
 function of \citet{Giallongo05}.  We convert their Schechter
 parameters to the $V$-band by correcting \lstar\ using the mean \bvr\
 in each redshift bin.  At $z=$0.73, 1.33, 2.01, and 2.8 our
 luminosity limit encompasses approximately 38, 40, 45, and 44\% of
 the total $V$-band light respectively.  We also check these numbers
 using the rest-frame $V$-band luminosity function from
 \citet{Marchesini06} and find that our \l{V} limit encompasses
 approximately 58 and 70\% of the light in our two highest redshift
 bins.  This is not to say that our relative trends in mass density
 will be more affected than other surveys that claim to measure
 ``total'' mass densities.  Some of these surveys, e.g. D03 and F03,
 assume a faint end slope of the luminosity function as determined at
 lower redshift and use it to extrapolate their observed quantities to
 fainter levels.  \rhostar\ is then derived by multiplying the
 resultant luminosity density by the mean \mlstar\ for their directly
 observed galaxies.  While formally integrating over all luminosities,
 the implicit assumption about the constancy of the faint end of the
 luminosity function is largely uncertain.  If the low end of the
 galaxy luminosity function is much steeper at high redshifts than at
 low redshifts this could cause the true decline rate in \rhostar\ to
 be less than what we observe.  In addition, if rest-frame optically
 faint galaxies have very different \mlstar\ values than brighter ones
 a bias in the determinations will occur.  The only solution to this
 quandary will be to go significantly deeper than the current deepest
 surveys in the \mk-band.  Given the substantial investment in
 telescope time required to obtain the HDF-S FIRES data (100 hours)
 only space-based observation will be able to push to significantly
 fainter limits.

 In R03 and in Figure~\ref{burstcol_fig} we demonstrated the
 advantages of averaging over the galaxy population when deriving
 stellar mass-to-light ratios.  One possible complication may result
 when the different populations being averaged have very different
 extinction properties, such as the difference between LBGs and DRGs
 in our sample.  We have explored the possible magnitude of such an
 error by fitting various combinations of models with SFHs and
 extinctions that correspond to star-forming DRGs, passive DRGs, and
 LBGs (e.g. P01; Shapley et al. 2001; F\"orster Schreiber et al. 2004;
 Labb\'e et al. 2005).  Specifically we have used 3 models consisting
 of a 1 Gyr old CSF model with $A_V\sim 2$, a 2.9 Gyr old passive
 population with no extinction, and a 300 Myr old galaxy with
 $A_V\sim0.6$.  Depending on the exact contributions of the different
 populations to the mean SED errors of up to a factor of 2 in the
 derived masses can exist.  We have also checked for the presence of
 these errors in our own data by using the DRG and non-DRG subsamples
 discussed in \S~\ref{massbudget_sec}.  On average these two
 subsamples have very different stellar populations and extinctions
 \citep{Forster04, Dokkum04, Labbe05} but the sum of the \rhostar\
 contributions from the two subsamples is within $15\%$ of the value
 measured for the total population, implying that this uncertainty is
 not dominant.

 We also have explored the effect of our limited choice of SFH on our
 results.  We increased the number of SFHs to include SSP, CSF,
 \taumod{100Myr}, \taumod{300Myr}, \taumod{500Myr}, \taumod{800Myr},
 \taumod{1Gyr}, \taumod{3Gyr}, \taumod{6Gyr}, and a SFH that
 corresponds to the globally averaged SFR($z$) from \citet{Cole01}.
 With this larger range in SFHs the best-fit \rhostar\ values decrease
 by $<15\%$ and the confidence limits in \rhostar\ remain unchanged.
 Therefore our choice of SFH is not a large source of error.

 Field-to-field variations are an obvious source of error, especially
 since the high luminosity galaxies we are examining are likely to be
 heavily clustered \citep{Adel04, Daddi03}.  We have attempted to
 mitigate this as much as possible by using data from every available
 field where suitably deep optical and NIR data are available.
 Obviously deep NIR imaging over many spatially disjoint fields is
 crucial to making progress in this arena.  This will require large
 investments of time on the next generation of wide-field NIR imagers
 on 8-meter class telescopes (e.g. HAWK-I/VLT, MOIRCS/Subaru).

 The IMF is poorly constrained at high redshift.  Although
 observations of the spectral signatures of massive stars in some LBGs
 indicate that the high mass slope must be close to Salpeter
 \citep{Pett00} some authors have argued for the necessity of a
 top-heavy IMF at high redshift to explain abundance ratios in
 elliptical galaxies (e.g. Matteucci 1994 and Nagashima et al. 2005)
 and the abundance of submillimeter galaxies at $z\sim2$
 \citep{Baugh05}.  As noted in \S~\ref{massevolve_sec} changing the
 lower end of the IMF would simply result in a scaling of all the
 \rhostar\ and SFR($z$) measurements by the same amount.  A
 non-universal IMF, however, either as a function of time,
 environment, or metallicity would result in a biased determination of
 the evolution in \rhostar\ and SFR($z$).

 In our modeling we assume solar metallicity and a \citet{Calz00}
 attenuation law.  It is impossible to constrain these directly using
 broad-band photometry.  To determine the sensitivity to these
 assumptions we re-fit our data using $Z=0.004{\rm Z_\odot}$ models
 and found that \rhostar\ declines by $0.1-0.25$dex.  This is somewhat
 less than the $0.3-0.5$dex change found by P01 and the reason is not
 entirely clear.  It may be because our mean SEDs are significantly
 redder at all redshifts than the very blue starburst SEDs modeled by
 P01.  We explore the dependence on the dust attenuation law by
 fitting using the SMC extinction curve of \citet{Prevot84}.  In this
 case \rhostar\ decreases by $<0.1$dex.  We conclude that uncertainties
 in the metallicity and dust extinction can result in up to a 0.2 dex
 error in our \rhostar\ estimates.

 We rely entirely on BC03 models to interpret the average SEDs.
 \citet{vanderwel06}, however, have recently pointed out that the BC03
 models incorrectly predict the evolution in \mlstarlam{K} and
 rest-frame $B-K$ color for early type galaxies at $z<1$.  The
 \citet{Maraston05} models do a better job at the fitting the
 evolution.  In the rest-frame optical, the BC03 and the Maraston
 models yield identical \mlstarlam{V} values for all models with
 $(U-V)_{rest}>0.5$ but the Maraston models yield \mlstarlam{V} values
 between 5 and 35\% higher for all models with $0<(U-V)_{rest}<0.5$
 (for solar metallicity and a Salpeter IMF).  Since our \uvr\ colors
 range from \uvr$\approx0$ at our highest redshift bin to \uvr$\approx
 1$ at $z=0$, adopting the Maraston models would make our mass density
 evolution shallower by 5-35\%.  This is not a dominant source of
 uncertainty in our analysis.

 Our analysis relies heavily on the use of photometric redshifts.  An
 incorrect estimate of the redshift will lead to errors in the
 rest-frame colors and luminosities, and hence in the stellar masses.
 Unfortunately the most thorough calibrations for \zp\ estimates have
 only been performed for UV-bright populations for which abundant
 spectroscopic redshifts are already in hand.  As we indicated in the
 previous section, however, much of the stellar mass density at high
 redshift likely resides in UV-faint objects, where \zp\ estimates are
 largely untested.  Nonetheless, initial progress with optical and NIR
 spectroscopy of DRGs (Kriek et al. 2006; Wuyts et al. in preparation)
 has shown that the photometric redshifts for these objects, while
 slightly less reliable than for optically selected samples, are
 rarely catastrophically wrong.  Looking at the whole spectroscopic
 sample with \l{V}$>$\lthreshlam{V} we find that \zp\ errors result in
 systematic errors in \l{V} of $\approx 4\%$ and $1\%$ at $z<1.5$ and
 $z\geq1.5$ respectively.  The systematic errors in \uvr\ are also
 small, $\approx 5\%$ and $11\%$, at $z<1.5$ and $z\geq1.5$
 respectively.  With the caveat that the spectroscopic samples are not
 representative of the NIR selected galaxy population we conclude that
 the photometric redshift errors should not be a dominant uncertainty
 in our analysis, although they are still important for the DRG
 population.

 A final concern stems from our use of a luminosity as opposed to a
 stellar mass cut to define our galaxy sample.  Galaxy luminosities
 can evolve rapidly due to bursts of star formation and rapidly
 changing dust contents.  For this reason galaxies may pop in and out
 of a luminosity selected sample as they evolve, implying that our
 sample may not be representative of the population as a whole.  In
 contrast to the luminosity, \mstar\ of a galaxy should evolve more
 slowly and a stellar mass selected sample will be less susceptible to
 the above problems.  This effect may not be significant, since our
 redshift bins are large enough in time to span many typical starburst
 timescales ($\lesssim$ a few 100 Myr) and the effect of galaxies
 entering and leaving our luminosity cut may average out.  To test the
 dependence of our results on the exact nature of our luminosity cut
 we have repeated all measurements using a passively evolving
 luminosity threshold corresponding to a maximally old population that
 has been normalized to have \l{V}$=$\lthreshlam{V} at $z=3.2$.  In
 the absence of dust evolution this cut ensures that all galaxies in
 our high redshift bins will also be present in the lower redshift
 bins, although galaxies which brightened or stay at constant
 luminosity as a function of time may enter the sample at lower
 redshifts.  With this evolving threshold the main effect is to
 increase \jrest\ and lower \mlstar\ as one moves to lower redshifts.
 These changes conspire lower $\rho_*(z=2.8)/\rho_*(z=0)$ by a factor
 of 1.9 compared to what is shown in Figures~\ref{massdens_fig} and
 \ref{masssim_fig} and in Tables~\ref{masstababs} and \ref{masstab}.

 We also estimate the change in the inferred evolution by evolving our
 \lthreshlam{V} to account for the observed change in \mlstarlam{V}.
 This translates into a 10.7 times fainter luminosity cut at $z\sim0$
 than at $z\sim 2.8$.  Using this fainter threshold, the resultant
 change in \rhostar\ out to $z\sim 2.8$ increases by a factor of 2.1.
 This would bring our measurements further out of agreement with the
 integral of the SFR($z$) curve.

 As a related concern, the use of a luminosity limited sample also
 complicates the comparison to other authors in
 \S~\ref{massevolve_sec}.  Our measured mean \mlstarlam{V} values and
 our \lthreshlam{V} limit at all redshifts correspond to a stellar
 mass that is significantly below our mass completeness
 limit\footnote{The mass completeness limit that corresponds to our
 \lthreshlam{V} is derived at each redshift using the \mlstarlam{V} of
 a maximally old single burst with a Salpeter IMF.}.  This implies
 that we are likely missing galaxies with high \mlstarlam{V} values.

 In summary, systematic errors in the derived \rhostar\ evolution may
 be present at the factor of $\sim 2$ level, with the dominant sources
 of error coming from our lack of knowledge about the faint end of the
 luminosity/mass functions and our use of a cut in rest-frame
 luminosity instead of mass.

\section{Discussion}
\label{discuss_sec}

\subsection{The Star Formation Rate Budget}

 As discussed in \S~\ref{massevolve_sec} and as shown in
 Figure~\ref{massdens_fig} the evolution in \rhostar\ roughly agrees
 with the integral of the SFR($z$), as calculated by \citet{Cole01}.
 This SFR was determined from rest-frame UV selected samples with a
 large (and uncertain) extinction correction applied.  That its
 integral agrees roughly with the direct \rhostar\ measurements
 indicates that the UV-selected samples trace most of the SFR density
 in the Universe, modulo the very large extinction corrections.  This
 is in agreement with recent work of \citet{Reddy05} and would in turn
 imply that the amount of heavily obscured star formation does not
 dominate the total, as long as the stars formed in those heavily
 obscured environments will later be visible in the rest-frame optical
 and hence enter into our \rhostar\ determinations.  \citet{Reddy05}
 also showed, however, that only $\lesssim 10\%$ of DRGs would be
 selected in traditional UV-selected samples.  Coupled with our
 measurement of the dominant DRG contribution to the mass budget at
 $z\geq2$ this indicates that UV-selected surveys are very incomplete
 in stellar mass.

\subsection{Constraining the Formation Epoch of Local Galaxies}

 It is interesting to ask what the observed evolution in \rhostar\
 implies for the formation times of different stellar populations in
 the local Universe.  Combining our measurements with the ``total''
 \rhostar\ measurements of other authors, the evolution in \rhostar\
 places the constraint that no more than 50\% of the stellar mass
 could have been formed at $z\gtrsim 1$. Indeed, various authors have
 found that $\sim 50\%$ of the local stellar mass in SDSS resides in
 early type galaxies \citep{Hogg02, Bell03, Kauff03}.  Age
 determinations of early type galaxies, both from stellar
 ``archeology'' (e.g. Trager et al. 2000; Thomas et al. 2005), and
 from fundamental plane studies at higher redshifts (e.g. van Dokkum
 et al. 1996; van der Wel et al. 2004) yield formation times for the
 stars in early type galaxies at $1<z<3$ in rough agreement with the
 evolution in \rhostar.  Likewise, the remainder of the local stellar
 mass density must have formed at $z\lesssim 1$ compatible with the
 expected formation times for stars in disks of the MW and M31 (e.g.
 Ferguson \& Johnson 2001; Hansen et al. 2002) and the inferred ages
 for large disk galaxies at $z<1$ \citep{Hammer05}.

\subsection{Comparison with Theoretical Predictions}

\begin{figure}
\ifemulate 
	\epsscale{1.2}
\else
	\epsscale{0.75}
\fi
\plotone{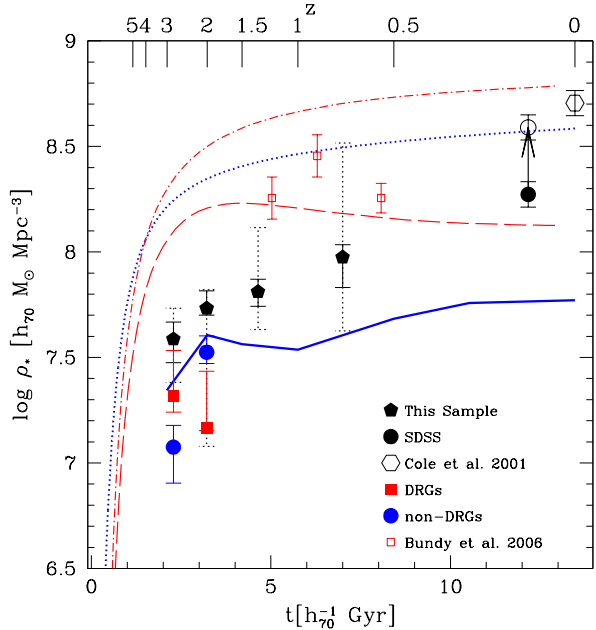}
\caption {A comparison of the evolution of of \rhostar\ with the predictions 
 of theoretical simulations.  The solid pentagons are our \rhostar\
 measurements for galaxies with \l{V}$>$\lthreshlam{V}, where the
 dotted errorbars indicate the minimum and maximum \rhostar\ values
 among our fields as an indication of the field-to-field variance.
 The solid circle is our measurement of \rhostar\ from SDSS for
 galaxies above the luminosity limit.  The open circle is our estimate
 of the total \rhostar\ from SDSS and the arrow connecting them
 indicates our correction to total \rhostar\ for the low redshift
 point.  The open hexagon is the total \rhostar\ measurement from
 \citet{Cole01}.  The red empty squares indicate the \rhostar\ values
 from \citet{Bundy06} derived for all galaxies with
 \l{V}$>$\lthreshlam{V}.  The dotted blue curve is the integral of the
 analytical SFR($z$) curve from \citet{Spring03}, which represents the
 total \rhostar\ from that simulation.  The solid blue curve is from a
 mock catalog created from the SH03 simulation \citep{Finlator06} and
 is for simulated galaxies with \l{V}$>$\lthreshlam{V}.  The results
 of the Millennium simulation \citep{Springel05, Croton05} are plotted
 as red curves.  The dot-dashed red curve corresponds to the total
 stellar mass density in the simulation and the dashed red curve
 corresponds to the mass density in simulated galaxies with
 \l{V}$>$\lthreshlam{V}.  The total \rhostar\ for all simulations
 agree with the observations at $z\sim0$.  When compared to the data
 using the same observational cut the \citet{Finlator06} simulations
 fail to match the ensemble of data at $z\lesssim1.4$.  The Millennium
 simulations, however, fail to match the observational data at
 $z\gtrsim1.4$.  We also show the \rhostar\ values for the non-DRGs
 and DRGs as filled blue circles and red squares respectively, in the
 two redshift bins where DRGs are selected.  }
\label{masssim_fig}
\end{figure}

 In Figure~\ref{masssim_fig} we show a comparison of our direct
 \rhostar\ measurements with the predictions for a set of theoretical
 models.  The thick dashed curve shows the evolution in \rhostar\
 derived by integrating an analytic fit to the SFR($z$) curve as
 computed from a set of nested hydrodynamic simulations that include
 the effects of star formation and feedback (Springel \& Hernquist
 2003; hereafter SH03).  This integral takes into account the mass
 loss from evolved stars.  The thin solid curve gives the prediction
 from a semi-analytic mock catalog of the Millennium Simulation
 (Springel et al. 2005; Croton et al. 2005; hereafter C05) that
 includes feedback from star formation and from active galactic nuclei
 (AGN). The total estimates of SH03 and C05 agree very well with each
 other, and with the local value of \rhostar.

 Because we compute our \rhostar\ measurements to a fixed \l{V} we can
 compare our data directly to theoretical models explicitly including
 the observational selection criteria.  We also include data from
 \citet{Bundy06} subjected to our same \l{V} limit (Bundy private
 communication).  Within the substantial field-to-field variations
 (dotted errorbars) our \rhostar\ estimates and those of
 \citet{Bundy06} are consistent.  \citet{Nagamine04} claim that the
 discrepancy between total \rhostar\ estimates and the model
 predictions is caused by observational selection effects.  According
 to those authors, the disagreement may be caused by the lack of
 observations at the faint end of the galaxy luminosity function
 coupled with the very steep galaxy stellar mass function in the
 simulations.  Because the simulations of \citet{Nagamine04} were not
 subjected to observational selection effects, however, it is
 difficult to ascertain the true nature of the disagreement.  We test
 this explicitly using mock catalogs subjected to our \lthreshlam{V}
 limit.  The dashed curve is the prediction of the Millennium
 simulation.  The solid curve is derived from a mock galaxy catalog
 computed from the SH03 simulation (Finlator et al. 2006; hereafter
 F06)\footnote{Specifically this mock catalog was computed with the G6
 simulation.}.  Both mock catalogs include the effect of extinction on
 the galaxy luminosities.  Contrary to the claim of \citet{Nagamine04}
 the disagreement between the models and the data at high redshift
 persists even when accounting for observational selection in the
 models.  The F06 mock catalog prediction has the same general slope
 as the SH03 total \rhostar\ estimate and although this model comes
 close to the data at high redshift it fails to match the ensemble of
 data at $z\lesssim 1.5$.  The evolution in \rhostar\ for luminous
 galaxies in the C05 catalog, however, agrees reasonably well with the
 data at $z\lesssim1.5$ but produces too much mass density in luminous
 galaxies at $z\lesssim1.5$.  In addition, it is worth noting that
 \rhostar\ of luminous galaxies in the Millennium simulation actually
 \textit{decreases} from $z\sim 2$ to $z\sim 0$.

 It is important to note that the overall normalization of the
 simulations is uncertain since the stellar mass depends intimately on
 the feedback, which is essentially a free parameter of the models,
 and can accommodate a 0.3 dex change in the all the curves.  The most
 striking disagreement with the models is the large difference in the
 shape of the \rhostar\ evolution compared to the observations,
 specifically in that the model evolution is too shallow, especially
 at $z<1.5$.

 It is difficult to disentangle the source of the disagreement between
 the models and the data.  That the model curves are too shallow
 implies that too much mass is formed into stars at high redshift,
 perhaps because of an incorrect cooling and feedback prescription.
 The F06 and SH03 models do not include AGN feedback which may modify
 the shape.  Any such feedback mechanism, however, must still be able
 to reproduce the observed number densities of massive galaxies at
 high redshift (e.g. van Dokkum et al. 2006).  The slightly decreasing
 \rhostar\ of luminous galaxies with cosmic time shown by the C05
 model at $z\lesssim 2$ indicates that the model galaxies are evolving
 rapidly in \mlstar.  This cannot be due to the mass loss from evolved
 stars since the C05 simulations instantaneously remove 30\% of the
 stellar mass as soon as that mass is formed.  This is an adequate
 approximation since 25\% of the stellar mass is already lost after 1
 Gyr.  The most likely explanation for this decline is therefore
 because objects that are above our luminosity cut at high redshift
 fade below it at lower redshifts.  The AGN feedback prescription
 adopted by the C05 models is the likely cause of this trend as it
 shuts off star formation in massive galaxies, subsequent to which
 they evolve passively to lower redshifts, with the commensurate
 amount of fading.  It is puzzling that the models have traditionally
 been faulted for not producing the proper number of massive galaxies
 at high redshift but now appear, at least for the C05 models, to
 produce too much mass in luminous galaxies at high redshift.

\subsection{Room for Improvement}

 There are various ways in which these measurements of \rhostar\ could
 be improved, each relating to specific sources of possible error:

 $\bullet$ Field-to-field variations are still a major source of
 uncertainty and can best be addressed by the inclusion of many more
 fields with deep optical and NIR imaging data, preferably to
 $K\sim23$.

 $\bullet$ We are still limited in that we are using a rest-frame
 luminosity selected sample.  With the deep IRAC data that are now
 becoming available more reliable and well constrained mass
 determinations may be possible for individual galaxies.  Selection by
 rest-frame \Ks-band will not entirely solve this problem.  In
 addition to the model uncertainties pointed out in
 \S\ref{syserr_subsec}, \citet{Labbe05} and \citet{Shap05} have
 demonstrated that there are still large (a factor of 6-13) variations
 in \mlstarlam{K} for high redshift galaxies and \citet{vanderwel06}
 have shown that \mlstarlam{K} of early type galaxies evolves by a
 factor of $\sim 3$ over $0<z<1$.  An IRAC-selected sample will not be
 the same as a stellar mass-selected sample.

 $\bullet$ There are few tests of photometric redshifts for galaxies
 with very red optical-UV colors.  Since it is impossible to measure
 spectroscopic redshifts for every object it is crucial to test
 photometric redshifts with extensive NIR spectroscopy performed with
 the new generation of multi-object NIR spectrographs.  In addition,
 these observations can give us independent measures of the dynamical
 masses of galaxies and will provide additional constraints on the
 \mstar\ estimates.

 $\bullet$ The uncertainty at the faint end of the luminosity/mass
 function is great.  Very deep NIR and MIR observations will be needed
 to see if the evolution inferred from the rest-frame optically bright
 population can be extended to the population as a whole.

 $\bullet$ Local studies are still inconclusive as to the existence of
 a universal IMF in the local Universe.  It is not clear at this time
 how to directly constrain the intermediate mass IMF at high
 redshifts.  Dynamical mass estimates or gravitational lensing
 measurements of galaxies in regions where the baryons are thought to
 dominate the total mass may provide some assistance.

\section{Summary and Conclusions}
\label{sum_sec}

 In this paper we have presented the evolution in the properties of
 the volume averaged stellar population from $z=0$ to $z=3$.  We have
 used deep NIR and optical imaging over $\sim 100$ square arcminutes
 in four disjoint fields to mitigate field-to-field variations.  The
 deep NIR data allow us to select and analyze galaxies at rest-frame
 optical wavelengths, where dust extinction and stellar mass-to-light
 ratio variations are far less than in the rest-frame UV.  We derived
 photometric redshifts for all objects lacking a spectroscopic
 redshift and measure rest-frame UV through optical luminosities of
 all galaxies at $z<3.2$.  We create a rest-frame \mv-band luminosity
 selected sample with \l{V}$>3\times 10^{10}$\lumsol\ above which we
 are complete at all redshifts.

 From these individual measurements we construct volume averaged
 properties such as the luminosity density, \jrest, and color,
 e.g. \ubr\ and \bvr.  The redshift evolution in luminosity density and
 mean color is very different.  As found in R03 for a much smaller
 sample, \jrestlam{V} is roughly constant within a factor of $\approx
 2.2$ from $z\sim3$ to $z=0$.  This modest evolution is broadly
 consistent with the results of D03 and \citet{Giallongo05}.  The
 evolution in the rest-frame colors, however, is dramatic and the
 colors become systematically redder with decreasing redshift.  The
 colors of our SDSS comparison sample at $z=0.1$ are \bvr$=0.72$ and
 \ubr$=0.07$.  In contrast the average color of luminous galaxies at
 $z\sim3$ is \bvr$=0.40$ and \ubr$=-0.38$.

 By averaging over the population as a whole we are also averaging
 over the individual SFHs of the galaxies and the resultant volume
 averaged SFH will be smoother than that of the individual galaxies.
 We expect the mean colors to be more easily modeled than those of
 individual galaxies, which presumably have bursty SFHs.  We find that
 the mean colors at all redshifts lie close to the locus of
 morphologically normal local galaxies in \bvr\ vs. \ubr\ space but
 compared to the local sample are slightly bluer in \ubr\ at a fixed
 \bvr.  Some of this difference could result from photometric redshift
 errors but could also be indicative of a residual signature of bursts
 in the mean colors.

 We derive the luminosity weighted volume averaged UV-optical
 rest-frame SED for all galaxies with \l{V}$>3\times 10^{10}$\lumsol.
 At all wavelengths the mean SED becomes gradually more red with
 decreasing redshift and at all redshifts is within the range of SEDs
 spanned by normal galaxies in the local Universe.  Even at our
 highest redshift bin the volume average SED does not look like that
 of a starburst from \citet{Kinn96} with little extinction but rather,
 is redder than an irregular template.  We also compute the mean SED
 from DRGs chosen to have \jk$>2.3$.  This mean DRG SED redder than
 average in the rest-frame optical and in the rest-frame UV.

 With simple models we transform the mean SED into an evolution in the
 global \mlstarlam{V}.  Because the volume averaged SFH should be
 smoother than that of the individual galaxies it is therefore more
 appropriate to use smooth models to fit the mean SED than to fit the
 SEDs of individual galaxies.  Nonetheless, the colors cannot be fit
 well at all redshifts with a single model having an exponential SFH
 and a constant dust obscuration and metallicity.  We can, however,
 fit the average SED at each redshift reasonably well with some
 combination of the parameters, as long as they can change as a
 function of redshift.  We use these model fits to derive luminosity
 weighted stellar mass-to-light ratios for all galaxies with
 \l{V}$>3\times 10^{10}$\lumsol.  \mlstarlam{V} also declines smoothly
 from $z=0.1$ to $z\sim3$ with \mlstarlam{V}$=4.9$ and $0.5$ at
 $z=0.1$ and $z\sim3$ respectively.  There are strong variations in
 \mlstarlam{V} across the population.  The DRGs have an \mlstarlam{V}
 a factor of $\approx 1.1$ and 3.6 higher than the non-DRGs in the
 $z=2.01$ and $z=2.8$ redshift bins.  This is consistent with the
 results of \citet{Labbe05} and \citet{Forster04} who modeled the SEDs
 of individual DRGs.

 Multiplying the \jrestlam{V} and mean \mlstarlam{V} measurements we
 derive the evolution in \rhostar\ out to $z\sim3$.  \rhostar\ in
 luminous galaxies declines by a factor of $4.8$ out to $z=3$.  At
 $z\sim 2$ and 2.8, the DRGs contribute 30\% and 64\% of \rhostar,
 showing that UV selection techniques miss much of the stellar mass
 at $1.6<z<3$ in rest-frame optically luminous galaxies.

 We also compare our \rhostar\ measurements to a set of recent model
 predictions subjected to our same rest-frame luminosity cut and find
 a large disagreement between the models themselves and between the
 models and observations.  The different models can match the observed
 evolution in \rhostar\ over a limited redshift range, but both models
 predict an evolution that is too shallow to match the observations.
 This therefore implies that the masses of luminous galaxies in the
 models, or their abundance, are not consistent with the observations.

 To better understand how different types of galaxies contribute to
 the stellar mass budget at high redshifts it will be necessary to
 reduce our reliance on photometric redshifts by performing NIR
 spectroscopy on a representative sample of DRGs.  This will confirm
 the redshifts of the optically faint population that contributes over
 50\% of the stellar mass at high redshift.  Re-performing this
 experiment with IRAC data will reduce the uncertainty due to dust
 obscuration and obtaining very deep NIR/MIR imaging will allow us to
 constrain the low mass end of the galaxy stellar mass function.

\acknowledgments

 We would like to thank Romeel Dav\'{e} for useful discussions, Darren
 Croton, Gerard Lemson, Gabriella De Lucia, and Jeremy Blaizot for
 supplying the output of the Millennium simulations, Christian Wolf
 for providing additional COMBO-17 data products, Michael Blanton for
 the creation of the SDSS data products, Henk Hoekstra for supplying
 us with his lensing corrections for MS1054-03, Kevin Bundy for
 providing us with mass density measurements for rest-frame \mv-band
 selected galaxies, and Danilo Marchesini for supplying us with a
 preliminary MUSYC rest-frame V-band luminosity function.  We would
 also like to thank our anonymous referee for their thoughtful and
 constructive comments.  GR would like to acknowledge the support of
 the Leo Goldberg Fellowship at the National Optical Astronomy
 Observatory.  The authors would also like to acknowledge the
 financial support of the Lorentz Center and the Leids
 Kerkhoven-Bosscha Fonds.

 Funding for the creation and distribution of the SDSS Archive has
 been provided by the Alfred P. Sloan Foundation, the Participating
 Institutions, the National Aeronautics and Space Administration, the
 National Science Foundation, the U.S. Department of Energy, the
 Japanese Monbukagakusho, and the Max Planck Society. The SDSS Web
 site is http://www.sdss.org/.

 The SDSS is managed by the Astrophysical Research Consortium (ARC)
 for the Participating Institutions. The Participating Institutions
 are The University of Chicago, Fermilab, the Institute for Advanced
 Study, the Japan Participation Group, The Johns Hopkins University,
 Los Alamos National Laboratory, the Max-Planck-Institute for
 Astronomy (MPIA), the Max-Planck-Institute for Astrophysics (MPA),
 New Mexico State University, University of Pittsburgh, Princeton
 University, the United States Naval Observatory, and the University
 of Washington.

\ifemulate
	\begin{deluxetable*}{rrrrrrrrr}
\else
	\begin{deluxetable}{rrrrrrrrr}
\fi
\tablecaption{Ultra-violet Rest-frame Luminosity Densities}
\tablewidth{0pt}
\tablehead{\colhead{$z$} & \colhead{${\rm Field(s)}$} & \colhead{$N_{tot}$} & \colhead{$N_{use}$} & \colhead{$j_{2200}$} & \colhead{$dj_{2200}$} & \colhead{$j_{2700}$} & \colhead{$dj_{2700}$}\\
\colhead{$(1)$} & \colhead{$(2)$} & \colhead{$(3)$} & \colhead{$(4)$} & \colhead{$(5)$} & \colhead{$(6)$} & \colhead{$(7)$} & \colhead{$(8)$}}
\startdata
$0.73$ & ALL & 116 & 112 & $-11.84$ & $0.34$ & $-12.61$ & $0.14$ \\
$1.33$ & ALL & 154 & 127 & $-12.35$ & $0.18$ & $-12.68$ & $0.14$ \\
$2.01$ & ALL & 164 & 126 & $-11.95$ & $0.14$ & $-12.38$ & $0.13$ \\
$2.80$ & ALL & 95 & 82 & $-13.67$ & $0.15$ & $-13.84$ & $0.14$ \\
$0.73$ & HDF-N & 11 & 11 & $-12.06$ & $0.42$ & $-12.79$ & $0.41$ \\
$1.33$ & HDF-N & 9 & 9 & $-13.15$ & $0.55$ & $-13.46$ & $0.50$ \\
$2.01$ & HDF-N & 13 & 11 & $-12.91$ & $0.38$ & $-13.18$ & $0.37$ \\
$2.80$ & HDF-N & 16 & 16 & $-13.82$ & $0.30$ & $-14.02$ & $0.30$ \\
$0.73$ & HDF-S & 4 & 4 & $-11.55$ & $0.58$ & $-12.12$ & $0.60$ \\
$1.33$ & HDF-S & 13 & 11 & $-12.78$ & $0.33$ & $-13.22$ & $0.33$ \\
$2.01$ & HDF-S & 7 & 6 & $-12.55$ & $0.51$ & $-12.84$ & $0.50$ \\
$2.80$ & HDF-S & 25 & 24 & $-14.60$ & $0.25$ & $-14.73$ & $0.24$ \\
$1.33$ & MS1054-03 & 36 & 34 & $-12.64$ & $0.24$ & $-12.90$ & $0.22$ \\
$2.01$ & MS1054-03 & 50 & 46 & $-12.35$ & $0.25$ & $-12.77$ & $0.24$ \\
$2.80$ & MS1054-03 & 54 & 42 & $-13.27$ & $0.22$ & $-13.46$ & $0.20$ \\
$0.73$ & CDF-S & 101 & 97 & $ $ & $ $ & $-12.62$ & $0.15$ \\
$1.33$ & CDF-S & 96 & 73 & $-12.07$ & $0.29$ & $-12.44$ & $0.22$ \\
$2.01$ & CDF-S & 94 & 63 & $-11.54$ & $0.20$ & $-12.04$ & $0.18$ \\
\enddata
\label{tablum_UV}
\tablecomments{All luminosity densities are given in $h_{70}$ AB magnitudes $/$ Mpc$^3$.  (3) is the total number of galaxies in each bin with \l{V}$>$\lthreshlam{V}.  In (4) we give the number of galaxies above our luminosity threshold that also have $\delta z / (1 + z) \leq 0.16$.  As described in \S\ref{ldens_meas_sec} and R03 we calculate the \jrest values using the galaxies in (4), and correct for those galaxies that have been excluded due to their large photometric redshift uncertainties.  The $z=0.73$ entry for MS1054-03 has been omitted because of the presence of a rich $z=0.83$ cluster in that field.  The $z=2.80$ entry for the CDF-S has been omitted because of a high incompleteness at these redshifts caused by the moderate depth.  The empty entry in columns (5) and (6) for the CDF-S is because $\lambda_{rest}=2200{\rm \AA}$ was blueward of the bluest observed filter.}

\ifemulate
	\end{deluxetable*}
\else
	\end{deluxetable}
\fi

\ifemulate
	\begin{deluxetable*}{rrrrrrrr}
\else
	\begin{deluxetable}{rrrrrrrr}
\fi
\tablecaption{Optical Rest-frame Luminosity Densities}
\tablewidth{0pt}
\tablehead{\colhead{$z$} & \colhead{${\rm Field(s)}$} & \colhead{$j_{U}$} & \colhead{$dj_{U}$} & \colhead{$j_{B}$} & \colhead{$dj_{B}$} & \colhead{$j_{V}$} & \colhead{$dj_{V}$}\\
\colhead{$(1)$} & \colhead{$(2)$} & \colhead{$(3)$} & \colhead{$(4)$} & \colhead{$(5)$} & \colhead{$(6)$} & \colhead{$(7)$} & \colhead{$(8)$}}
\startdata
$0.73$ & ALL & $-13.86$ & $0.12$ & $-14.82$ & $0.12$ & $-15.40$ & $0.12$ \\
$1.33$ & ALL & $-13.73$ & $0.09$ & $-14.52$ & $0.09$ & $-14.98$ & $0.09$ \\
$2.01$ & ALL & $-13.53$ & $0.09$ & $-14.31$ & $0.09$ & $-14.78$ & $0.08$ \\
$2.80$ & ALL & $-14.55$ & $0.11$ & $-15.05$ & $0.12$ & $-15.35$ & $0.12$ \\
$0.73$ & HDF-N & $-14.12$ & $0.40$ & $-15.15$ & $0.38$ & $-15.73$ & $0.37$ \\
$1.33$ & HDF-N & $-14.26$ & $0.41$ & $-15.03$ & $0.39$ & $-15.52$ & $0.41$ \\
$2.01$ & HDF-N & $-14.01$ & $0.28$ & $-14.59$ & $0.28$ & $-14.93$ & $0.30$ \\
$2.80$ & HDF-N & $-14.48$ & $0.31$ & $-14.90$ & $0.29$ & $-15.20$ & $0.28$ \\
$0.73$ & HDF-S & $-13.21$ & $0.59$ & $-14.12$ & $0.61$ & $-14.64$ & $0.62$ \\
$1.33$ & HDF-S & $-14.36$ & $0.26$ & $-15.17$ & $0.27$ & $-15.62$ & $0.26$ \\
$2.01$ & HDF-S & $-13.63$ & $0.40$ & $-14.20$ & $0.38$ & $-14.46$ & $0.38$ \\
$2.80$ & HDF-S & $-15.22$ & $0.21$ & $-15.70$ & $0.19$ & $-15.93$ & $0.19$ \\
$1.33$ & MS1054-03 & $-13.72$ & $0.18$ & $-14.43$ & $0.17$ & $-14.85$ & $0.17$ \\
$2.01$ & MS1054-03 & $-13.76$ & $0.18$ & $-14.55$ & $0.17$ & $-15.00$ & $0.17$ \\
$2.80$ & MS1054-03 & $-14.35$ & $0.15$ & $-14.89$ & $0.16$ & $-15.21$ & $0.17$ \\
$0.73$ & CDF-S & $-13.87$ & $0.13$ & $-14.83$ & $0.13$ & $-15.40$ & $0.12$ \\
$1.33$ & CDF-S & $-13.62$ & $0.12$ & $-14.44$ & $0.11$ & $-14.91$ & $0.11$ \\
$2.01$ & CDF-S & $-13.37$ & $0.11$ & $-14.18$ & $0.11$ & $-14.69$ & $0.11$ \\
\enddata
\label{tablum_opt}
\tablecomments{All luminosity densities are given in $h_{70}$ AB magnitudes $/$ Mpc$^3$ with the following conversion to Vega magnitudes: $U_{vega} = U_{AB} - 0.79$, $B_{vega} = B_{AB} + 0.102$, $V_{vega} = V_{AB} - 0.008$.  The $z=0.73$ entry for MS1054-03 has been omitted because of the presence of a rich $z=0.83$ cluster in that field.  The $z=2.80$ entry for the CDF-S has been omitted because of a high incompleteness at these redshifts caused by the moderate depth.  }
\ifemulate
	\end{deluxetable*}
\else
	\end{deluxetable}
\fi

\ifemulate
	\begin{deluxetable*}{ccccccc}
\else
	\begin{deluxetable}{ccccccc}
\fi
\tablecaption{Stellar Mass Densities}
\tablewidth{0pt}
\tablehead{\colhead{$z$} & \colhead{${\rm L}_{V}^{\rm thresh}$} & \colhead{${\rm log}(\rho_{\star})$} & \colhead{${\rm log}(\rho_{\star,low})$} & \colhead{${\rm log}(\rho_{\star,high})$} & \colhead{${\rm log}(\rho_{\star,low,f2f})$} & \colhead{${\rm log}(\rho_{\star,high,f2f})$}\\
\colhead{$(1)$} & \colhead{$(2)$} & \colhead{$(3)$} & \colhead{$(4)$} & \colhead{$(5)$} & \colhead{$(6)$} & \colhead{$(7)$}}
\startdata
$0.1$ & 0  & $ 8.59 $ &$ 8.55$& $ 8.63 $ \\
$0.1$ & $3\times 10^{10}$ & $ 8.27 $&$ 8.23$  & $ 8.31 $ \\
$0.73$ & $3\times 10^{10}$ & $7.97$ & $7.83$ & $8.03$ & $ 7.63 $ & $ 8.52 $\\
$1.33$ & $3\times 10^{10}$ & $7.81$ & $7.74$ & $7.87$ & $ 7.63 $ & $ 8.12 $\\
$2.01$ & $3\times 10^{10}$ & $7.73$ & $7.70$ & $7.82$ & $ 7.08 $ & $ 7.82 $\\
$2.80$ & $3\times 10^{10}$ & $7.59$ & $7.47$ & $7.67$ & $ 7.38 $ & $ 7.73 $\\
$0.1$ & $0.3\times 10^{10}$ \tablenotemark{a} & $ 8.59 $&$ 8.55$  & $ 8.63 $ \\
$0.73$ & $1.3\times 10^{10}$ \tablenotemark{a}& $8.05$ & $8.02$ & $8.12$ & $ 7.86 $ & $ 8.52 $\\
$1.33$ & $1.3\times 10^{10}$ \tablenotemark{a}& $7.87$ & $7.83$ & $7.94$ & $ 7.74 $ & $ 8.22 $\\
$2.01$ & $1.3\times 10^{10}$ \tablenotemark{a}& $7.76$ & $7.70$ & $7.82$ & $ 7.31 $ & $ 7.89 $\\
$2.80$ & $3\times 10^{10}$ \tablenotemark{a}& $7.59$ & $7.48$ & $7.65$ & $ 7.38 $ & $ 7.73 $\\
\enddata
\label{masstababs}
\tablecomments{(2) is given in units of ${\rm h_{70}^{-2}~L_{V,\odot}}$. (3)--(7) are given in units of ${\rm h_{70}}~{\cal M_{\odot}}$ Mpc$^{-3}$ and are for all galaxies with \l{V}$>$\lthreshlam{V}. The first 2 rows indicate our measurements for the SDSS. (4) and (5) are the formal confidence limits on \rhostar. (6) and (7) give the range in \rhostar\ allowed by the rms field-to-field variations among our different fields.}
\tablenotetext{a}{These \l{V} limits are computed using the observed variation in \mlstarlam{V} for galaxies with \l{V}$> 3\times 10^{10}$${\rm h_{70}^{-2}~L_{V,\odot}}$.}
\ifemulate
	\end{deluxetable*}
\else
	\end{deluxetable}
\fi

\ifemulate
	\begin{deluxetable*}{ccccc}
\else
	\begin{deluxetable}{ccccc}
\fi
\tablecaption{Mean Mass-to-light ratios}
\tablewidth{0pt}
\tablehead{\colhead{$z$} & \colhead{${\rm Field(s)}$} & \colhead{$\langle M_\star/L_{V}\rangle$} & \colhead{$\langle {\cal M_\star}/L_{V}\rangle_{low}$} & \colhead{$\langle M_\star/L_{V}\rangle_{high}$}\\
$(1)$ & $(2)$ & $(3)$ & $(4)$ & $(5)$}
\startdata
$0.73$ & ALL & $1.06$ & $0.77$ & $1.18$ \\
$1.33$ & ALL & $1.07$ & $0.88$ & $1.18$ \\
$2.01$ & ALL & $1.08$ & $0.97$ & $1.19$ \\
$2.80$ & ALL & $0.46$ & $0.39$ & $0.50$ \\
$0.73$ & HDF-N & $1.30$ & $1.06$ & $1.44$ \\
$1.33$ & HDF-N & $0.97$ & $0.65$ & $1.18$ \\
$2.01$ & HDF-N & $0.58$ & $0.40$ & $0.76$ \\
$2.80$ & HDF-N & $0.33$ & $0.15$ & $0.39$ \\
$0.73$ & HDF-S & $0.96$ & $0.85$ & $1.36$ \\
$1.33$ & HDF-S & $1.19$ & $0.78$ & $1.26$ \\
$2.01$ & HDF-S & $0.32$ & $0.28$ & $0.62$ \\
$2.80$ & HDF-S & $0.37$ & $0.22$ & $0.43$ \\
$1.33$ & MS1054-03 & $0.80$ & $0.71$ & $0.97$ \\
$2.01$ & MS1054-03 & $1.08$ & $0.78$ & $1.19$ \\
$2.80$ & MS1054-03 & $0.50$ & $0.37$ & $0.62$ \\
$0.73$ & CDF-S & $1.06$ & $0.96$ & $1.06$ \\
$1.33$ & CDF-S & $1.08$ & $0.78$ & $1.18$ \\
$2.01$ & CDF-S & $1.26$ & $1.03$ & $1.39$ \\
\enddata
\label{mltab}
\tablecomments{Columns (3)-(5) are computed for all galaxies with \l{V}$>$\lthreshlam{V} and are in units of $[{\cal M_{\odot}}/L_{\odot}]$.  The $z=0.73$ entry for MS1054-03 has been omitted because of the presence of a rich $z=0.83$ cluster in that field.  The $z=2.80$ entry for the CDF-S has been omitted because of a high incompleteness at these redshifts caused by the moderate depth.  }
\ifemulate
	\end{deluxetable*}
\else
	\end{deluxetable}
\fi

\ifemulate
	\begin{deluxetable*}{ccccccc}
\else
	\begin{deluxetable}{ccccccc}
\fi
\tablecaption{Relative Stellar Mass Densities}
\tablewidth{0pt}
\tablehead{\colhead{$z$} & \colhead{${\rm Field(s)}$} & \colhead{$\Delta\rho_{\star}$} & \colhead{$\Delta\rho_{\star,low}$} & \colhead{$\Delta\rho_{\star,high}$} & \colhead{$\Delta\rho_{\star,low}^{corr}$} & \colhead{$\Delta\rho_{\star,high}^{corr}$}\\
$(1)$ & $(2)$ & $(3)$ & $(4)$ & $(5)$ & $(6)$ & $(7)$}
\startdata
$0.73$ & ALL & $0.39$ & $0.28$ & $0.44$ & $0.21$ & $0.44$ \\
$1.33$ & ALL & $0.27$ & $0.23$ & $0.30$ & $0.13$ & $0.30$ \\
$2.01$ & ALL & $0.22$ & $0.21$ & $0.27$ & $0.10$ & $0.27$ \\
$2.80$ & ALL & $0.16$ & $0.12$ & $0.19$ & $0.06$ & $0.19$ \\
$0.73$ & HDF-N & $0.64$ & $0.38$ & $0.83$ & $0.18$ & $0.83$ \\
$1.33$ & HDF-N & $0.39$ & $0.19$ & $0.57$ & $0.09$ & $0.57$ \\
$2.01$ & HDF-N & $0.14$ & $0.08$ & $0.22$ & $0.04$ & $0.22$ \\
$2.80$ & HDF-N & $0.10$ & $0.05$ & $0.13$ & $0.02$ & $0.13$ \\
$0.73$ & HDF-S & $0.17$ & $0.07$ & $0.32$ & $0.04$ & $0.32$ \\
$1.33$ & HDF-S & $0.53$ & $0.33$ & $0.69$ & $0.16$ & $0.69$ \\
$2.01$ & HDF-S & $0.05$ & $0.04$ & $0.10$ & $0.02$ & $0.10$ \\
$2.80$ & HDF-S & $0.22$ & $0.13$ & $0.28$ & $0.06$ & $0.28$ \\
$1.33$ & MS1054-03 & $0.18$ & $0.14$ & $0.22$ & $0.07$ & $0.22$ \\
$2.01$ & MS1054-03 & $0.27$ & $0.19$ & $0.33$ & $0.09$ & $0.33$ \\
$2.80$ & MS1054-03 & $0.15$ & $0.10$ & $0.21$ & $0.05$ & $0.21$ \\
$0.73$ & CDF-S & $0.39$ & $0.34$ & $0.44$ & $0.16$ & $0.44$ \\
$1.33$ & CDF-S & $0.25$ & $0.19$ & $0.29$ & $0.09$ & $0.29$ \\
$2.01$ & CDF-S & $0.24$ & $0.20$ & $0.29$ & $0.10$ & $0.29$ \\
\enddata
\label{masstab}
\tablecomments{The relative mass densities in (3) are given in terms of the fractional decrease from $z=0$, i.e. $\Delta\rho_\star = \rho_\star/\rho_\star(z=0)$, as computed by correcting the values to total using the SDSS data and using the $z=0$ measurement of \citet{Cole01}.  The confidence intervals in $\Delta\rho_\star$ given in columns (4) and (5) come from the formal uncertainties on this evolution for galaxies with \l{V}$>$\lthreshlam{V}.  The confidence intervals in columns (6) and (7) include the extra uncertainty that comes from allowing \lthreshlam{V} to evolve with redshift according to the evolution in \mlstarlam{V}.  The $z=0.73$ entry for MS1054-03 has been omitted because of the presence of a rich $z=0.83$ cluster in that field.  The $z=2.80$ entry for the CDF-S has been omitted because of a high incompleteness at these redshifts caused by the moderate depth.  }
\ifemulate
	\end{deluxetable*}
\else
	\end{deluxetable}
\fi

\clearpage

\appendix
\section{The Effect of Weak Lensing on the Luminosity Density}
\label{lum_dens_lens_app}

 Weak lensing of background galaxies by a foreground mass distribution
 affects the luminosity density of the background sources by changing
 their apparent magnitudes, surface densities, and the volume enclosed
 in the observed solid angle.  To the first order all of these effects
 are directly dependent on the magnification factor $\mu$.  Given a
 mass map for the foreground cluster and the redshift of each source
 it is possible to compute the magnifications of each source $\mu_i$.

 First we consider the effect of weak lensing on the total integrated
 luminosity density in a given redshift slice whose enclosed co-moving
 volume per unit solid angle is $V_0$, in units of $[{\rm Mpc^3 /
 ster}]$.  The ``true'' luminosity density, i.e. in the absence of
 lensing, is $j_o$.  This can be written as:
\begin{equation}
j_0 = \frac{n_0}{V_0} \sum_{i=0}^{n_{gal}} \frac{L_{0, i}}{n_{gal}}~\left[\frac{\rm L_{\odot}}{\rm Mpc^3}\right],
\end{equation}

 where $n_0$ is the surface density of sources in units of $[{\rm
 ster}^{-1}]$.  $n_0$ can be written as

\begin{equation}
n_0 = N_0 V_0~[{\rm ster}^{-1}],
\end{equation}

 where $N_0$ is the intrinsic volume number density in units of $[{\rm
 \# / Mpc^3}]$.  We can now rewrite $j_0$ as

\begin{equation}
j_0 = N_0 \sum_{i=0}^{n_{gal}} \frac{L_{0, i}}{n_{gal}} ~\left[{\rm L_{\odot} / Mpc^3}\right].
\end{equation}

Under the influence of lensing the rest-frame luminosities (which is
what we derive from observed magnitudes) become

\begin{equation}
L_{\ell,i} = L_{0,i} \mu_i
\end{equation}

while the intrinsic volume number density naturally remains
unaffected.  The luminosity density subject to lensing $j_{\ell}$ is
therefore

\begin{equation}
j_\ell = N_0 \sum_{i=0}^{n_{gal}} \frac{L_{0, i} \mu_i}{n_{gal}} = \frac{n_0}{V_0} \sum_{i=0}^{n_{gal}} \frac{L_{0, i} \mu_i}{n_{gal}}~\left[{\rm L_{\odot} / Mpc^3}\right].
\end{equation}

 Since the sole difference between $j_\ell$ and $j_0$ is the
 magnification of the individual sources, we compute $j_0$ in
 MS1054-03 by using the demagnified galaxy luminosities, $L_{0, i}=
 L_{\ell, i} / \mu_i$.  The $\mu_i$ are determined from the weak
 lensing map of \citet{Hoekstra00}.  

 The second effect stems from the use of a fixed luminosity limit in
 the presence of magnification.  For example, sources whose intrinsic
 luminosities would exclude them from the sample can enter the sample
 after being magnified.  At all but the highest redshifts in the
 MS1054-03 field our data are complete significantly below the
 luminosity limit.  Therefore to correct for this effect in our
 analysis we select galaxies brighter than \lthreshlam{V} based on
 their demagnified luminosities.

\end{document}